\documentclass[aps,amsmath,prd,preprint,floats,epsf,superscriptaddress,nofootinbib]{revtex4}
\usepackage{graphicx}
\usepackage{bm}
\usepackage{color}

\def\gev{{\rm GeV}}
\def\tev{{\rm TeV}}

\begin{document}

\preprint{IPMU 18-0116}
\bigskip

\title{Muon $g-2$ and rare top decays in up-type specific \\ variant axion models}

\author{Cheng-Wei Chiang}
\email[e-mail: ]{chengwei@phys.ntu.edu.tw}
\affiliation{Department of Physics, National Taiwan University, Taipei, Taiwan 10617, R.O.C.}
\affiliation{Institute of Physics, Academia Sinica, Taipei, Taiwan 11529, R.O.C.}
\affiliation{Kavli IPMU (WPI), UTIAS, University of Tokyo, Kashiwa, Chiba 277-8583, Japan}

\author{Michihisa Takeuchi}
\email[e-mail: ]{michihisa.takeuchi@ipmu.jp}
\affiliation{Kavli IPMU (WPI), UTIAS, University of Tokyo, Kashiwa, Chiba 277-8583, Japan}

\author{Po-Yan Tseng}
\email[e-mail: ]{poyen.tseng@ipmu.jp}
\affiliation{Kavli IPMU (WPI), UTIAS, University of Tokyo, Kashiwa, Chiba 277-8583, Japan}

\author{Tsutomu T. Yanagida}
\email[e-mail: ]{tsutomu.tyanagida@ipmu.jp}
\affiliation{Kavli IPMU (WPI), UTIAS, University of Tokyo, Kashiwa, Chiba 277-8583, Japan}

\date{\today}

\begin{abstract}
The invisible variant axion models (VAM's) offer a very attractive solution for the strong CP problem without the domain wall problem. 
We consider the up-type specific variant axion models and examine their compatibility with the muon $g-2$ anomaly and the constraints from lepton flavor universality, several flavor observables, 
and top quark measurements.
We find that the combined $\chi^2$ fit favors the parameters $m_A\sim 15~\gev$ and $\tan\beta \sim 40$, 
the same as the type-X 2HDM. 
Moreover, we find that there are no conflict with any flavor observables as long as the mixing angle $\rho_u$ is sufficiently small.  In particular, a small nonzero mixing angle $\rho_u \sim \pi/100$ is slightly favored by the observed $B_s \to \mu\mu$ branching ratio.
The up-specific VAM predicts the flavor-violating top rare decay $t\to uA$ followed by $A \to \tau\tau$, which would provide a smoking gun signature at the LHC.  We show that current searches of $A$ already impose 
some constraints on the parameter space but are not sensitive to the most interesting light $m_A$ region.
We propose an efficient search strategy that employs di-tau tagging using jet substructure information, and 
demonstrate that it can enhance the sensitivity on $BR(t \to uA)$, especially in the light $m_A$ region. 
This model also predicts the flavor-violating decay of heavy Higgs bosons, such as $H \to t u$, that would suppress the $H \to \tau\tau/\mu\mu$ decays.  We also examine the up-specific VAM with the muon-specific lepton sector and the down-type specific VAM's as interesting alternative scenarios. 
\end{abstract}

\pacs{}

\maketitle

\section{Introduction}
\label{sec:intro}

The strong CP problem of an empirically tiny CP-violating phase in QCD, $\theta_{\rm QCD}$, can be solved by employing a $U(1)$ Peccei-Quinn (PQ) symmetry~\cite{Peccei:1977hh}, with which one can rotate away the undesired phase.  Such a $U(1)_{PQ}$ is assumed to be anomalous and broken spontaneously, resulting in the existence of a pseudo Nambu-Goldstone boson, called the axion~\cite{Weinberg:1977ma,Wilczek:1977pj}, whose dynamics is characterized by the axion decay constant $f_a$.  Such a model is subject to various experimental constraints.  Axion helioscopes and astronomical observations give a lower bound of $f_a \gtrsim 10^9$\,GeV (see, for example, Ref.~\cite{Agashe:2014kda}).  On the other hand, coherent oscillations of the axion field can play the role of a cold dark matter in the Universe~\cite{Abbott:1982af,Preskill:1982cy,Dine:1982ah}, from which one has $f_a \sim 10^{11-12}$~GeV~\cite{Ade:2015xua}, provided that the axion is the dominant component of the dark matter.  However, the model has a serious problem of domain wall formation in the early Universe.  This is because the number of discrete vacua separated by the domain walls is related to the number of fermion generations, which is 3 in the standard model (SM).  Such a problem can be resolved by assuming that only one right-handed (RH) quark is charged under the PQ symmetry, thus rendering a unique vacuum~\cite{Peccei:1986pn,Krauss:1986wx}.

Consistency of the axion model requires the use of two Higgs doublet fields $\Phi_1$ and $\Phi_2$, with $\Phi_1$ charged under the PQ symmetry while $\Phi_2$ being neutral.  Such an arrangement of assigning PQ charges to one Higgs doublet field and one RH quark would lead to flavor-changing neutral scalar (FCNS) couplings in the quark sector~\cite{Chen:2010su}.  Depending on whether the RH quark belongs to the up or down sector, the FCNS interactions could respectively happen among the up- or down-type quarks.  For example, a top-specific variant axion model (VAM) has recently been studied in Refs.~\cite{Chiang:2015cba,Chiang:2017fjr}.  Since such FCNS couplings depend on the chirality of fermions, the VAM presents a different Yukawa structure from, for example, the common two-Higgs doublet models (2HDM's).  As the SM lepton sector is irrelevant to the above-mentioned domain wall problem, one has the freedom of assigning either zero or non-zero PQ charge to the leptons.

In general, we have six possible choices of assigning a non-zero PQ charge to one of the RH quark fields: $u_R, c_R, t_R, d_R, s_R$ and $b_R$.  In this work, we mainly discuss a scenario in which one of the up-type RH quark fields is charged under $U(1)_{PQ}$.  Depending on the mixing parameters, one can obtain as a special case the top-specific VAM examined in Refs.~\cite{Chiang:2015cba,Chiang:2017fjr}.  
We will show that the mixing parameters are constrained 
under various experimental constraints such as the Higgs signal strength data and neutral $D$ meson mixing, 
and left with three possible regions corresponding to up-specific, charm-specific and top-specific scenarios. 
The scenario of having one of the down-type RH quark fields charged under $U(1)_{PQ}$, 
as we will show, is more severely constrained by low-energy flavor physics data.

Motivated by the long-standing puzzle of a $3\sigma$-level deviation~\cite{Hagiwara:2011af} in the muon anomalous magnetic dipole moment ($a_\mu$ or $(g-2)_\mu$) from the SM, we find it advantageous to make all leptons charged under the PQ symmetry as well in the model.  In fact, as far as the lepton sector and the third-generation quarks are concerned, the up-specific and charm-specific models become effectively identical to the usual Type-X 2HDM.  Since the up or charm Yukawa coupling is too small to affect the direct search constraints of additional Higgs bosons in collider experiments, which are mainly determined by the third-generation Yukawa couplings, most of the same constraints on Type-X 2HDM can be directly applied here.  In particular, it is known that Type-X 2HDM is difficult to constrain at the LHC and that it can explain the muon $g-2$ deviation by taking large $\tan\beta$ values of $\sim 40-50$ and a light pseudo-scalar Higgs boson $A$ with $m_A\sim 20~\gev$~\cite{2HDMg2}. One of the efforts to constrain such a model at the LHC can be found in Ref.~\cite{Chun:2015hsa}.  
We will see that the up-specific model shares the same parameter set to explain the $(g-2)_\mu$ deviation, 
while the charm-specific model is not favored as the two-loop contribution to $(g-2)_\mu$ is not negligible and has the opposite sign. 
Note that the requirements of large $\tan\beta$ and perturbativity of the top 
Yukawa coupling prohibit us from assigning a non-zero PQ charge to the RH top quark,
as examined in Ref.~\cite{Chiang:2017fjr}.

In the up/charm-specific model with the above-mentioned setup, an interesting rare top decay $t \to u/c A$ followed by $A \to \tau\tau$ or $\mu\mu$ is predicted.  Even though there is no dedicated experimental study focusing on this process, we find that searches for $bbA$ production followed by $A \to \tau\tau$ (or $\mu\mu$) already constrains the parameter space of this model.  Nevertheless, a dedicated search of the $t \to u/c A$ rare decay would still provide a better sensitivity to this model. We will propose an efficient strategy using the tau-tag algorithm with the jet substructure information and 
show that the sensitivity would be much enhanced.
In this model, the heavy Higgs bosons also have flavor-violating decay modes.  Those flavor-violating processes would provide smoking-gun signatures of the model at the LHC.

As a solution to the domain wall problem, we have more freedom in assigning the PQ charges in the lepton sector.  If only $\mu$ is PQ charged, the lepton sector becomes identical to the so-called muon-specific 2HDM, which is shown to successfully accommodate $(g-2)_\mu$ without relying on the 2-loop contribution with a light $A$ boson~\cite{Abe:2017jqo}.
We will see that the up-specific VAM with the muon-specific lepton sector is another attractive possibility as it is not constrained by the lepton universality measurements and no tuning is required to suppress $h \to AA$, thanks to the absence of such light particles.
Unlike the original muon-specific model, the up-specific VAM with the muon-specific sector 
predicts that the heavy Higgs bosons can decay into a pair of flavor-violating up-type quarks 
such as $H/A \to t u$ at a significant branching fraction.
It thus suppresses the $H/A \to \mu\mu$ decay, making the $4\mu$ constraint at the LHC 
less effective and opening up more parameter space.

This paper is organized as follows.  Section~\ref{sec:Lagrangian} discusses the structure of the Higgs sector 
in the VAM and possible scenarios of PQ charge assignment to the quark fields.  
We work out the FCNS couplings of the SM-like Higgs boson to the SM fermions.  
Section~\ref{sec:current_bounds} is devoted to studying possible effects of the 
FCNS interactions on physical observables.  Current data such as the muon $g-2$ anomaly, 
the lepton universality in $\tau$ decays, the lepton universality in $Z$ decays, rare $B$ decays 
and $D$ meson mixing, and top observables are imposed to constrain the mixing parameters in the Higgs sector.  In Section~\ref{sec:quark_FCNC}, we focus on a promising signature of the model, 
namely the rare $t \to u A$ and $t \to c A$ decays.  We show the current constraints from existing searches, and propose an effective way to look for the signature of this model by introducing boosted $A \to \tau\tau$ tagging using jet substructures.
Section~\ref{sec:down_type_variant} discuss another interesting possibility of the charge assignments for the lepton sector and 
how the down-type VAM is severely constrained by data.
We summarize our findings of this study in Section~\ref{sec:conclusion}.

\section{Up-Type Specific Variant Axion Model}
\label{sec:Lagrangian}

In a minimal setup of the VAM, we have two Higgs doublet fields $\Phi_1$ and $\Phi_2$ and 
a scalar field $\sigma$ with PQ charges $-1$, $0$ and $1$, respectively.
\footnote{Note that there is a difference in the convention between this work and Ref.~\cite{Chiang:2017fjr}. The PQ-charged Higgs field is $\Phi_1$ in the former case and $\Phi_2$ in the latter.}  
The $\sigma$ field, a SM gauge singlet scalar, is introduced to break the PQ symmetry spontaneously at a high energy scale by acquiring a vacuum expectation value (VEV) $f_a$ while it does not play much a role at low energies.  In general, one can choose any one of the six RH quark fields to be charged under $U(1)_{PQ}$.  It is also all right for leptons to carry either zero or non-zero PQ charges.  After electroweak symmetry breaking, $\Phi_{1,2}$ acquire the VEV's $v_{1,2}$, respectively.  Empirically, $v \equiv \sqrt{v_1^2 + v_2^2} = (246~{\rm GeV})^2$.  We define $\tan\beta = v_2 / v_1$ following the usual convention in the 2HDM's.

We first argue that the requirements of accommodating the muon $g-2$ anomaly and perturbativity in the Yukawa couplings largely restrict ourselves to only the scenario of up-type specific VAM's.  Consider first the scenario where the leptons do not carry the PQ charge and therefore have to couple only with $\Phi_2$.  In this case, the VEV's of the Higgs fields must satisfy the hierarchy: $\langle \Phi_2 \rangle \ll \langle \Phi_1 \rangle$.  In order to reproduce its mass, the RH top quark has to carry a non-zero PQ charge and therefore couple to $\Phi_1$.  As far as the down-type quark sector and the lepton sector are concerned, the model is identical to the Type-II 2HDM.  However, it had been shown that such a model could not explain $\Delta a_\mu$~\cite{2HDMg2}.
Therefore, both RH top and RH bottom quarks have to have non-zero PQ charges and couple to $\Phi_1$.  Nevertheless, such a PQ charge assignment gives rise to the domain wall problem.  As a conclusion, the scenario where the leptons do not carry the PQ charge fails to solve the muon $g-2$.

In the scenario where the leptons are charged under $U(1)_{PQ}$, the VEV's must satisfy instead $\langle \Phi_1 \rangle \ll \langle \Phi_2 \rangle$ (corresponding to $\tan\beta \gg 1$) in order for the lepton Yukawa couplings to be sufficiently large to explain $\Delta a_\mu$.  In this case, $t_R$ cannot be the one carrying a non-zero PQ charge because the top Yukawa would become non-perturbative.  We are then left with the choices of assigning a non-zero PQ charge to one of the remaining five RH quark fields.  In the following, we will formulate the up-type specific VAM's as an explicit example, and comment on stringent constraints on the down-type specific VAM's from low-energy flavor physics data.

Following the above argument, we assign a non-zero PQ charge of $-1$ to the RH up or charm quark field $u_R/c_R$.  As explicitly shown below, these two possibilities are related by a rotation in the field space.  The Yukawa interactions are given by:
\begin{eqnarray}
\label{eq3}
\mathcal{L} &=& 
-\Phi_1 \bar{u}_{R1}[Y_{u1}]_{i}Q_i
-\Phi_2 \bar{u}_{Ra}[Y_{u2}]_{ai}Q_i
-\Phi_1 \bar{e}_{Rj}[Y_{e}]_{ji}L_i
-\Phi_2 \bar{d}_{Rj}[Y_{d}]_{ji}Q_i + {\rm h.c.} ~, 
\end{eqnarray}
where the family indices $a \in \{ 2,3 \}$, and $i,j \in \{ 1,2,3 \}$.  Explicitly, $Y_{u1,u2}$ assume the forms of
\begin{align}
Y_{u1}=
\begin{pmatrix}*&*&*\cr 
0 &0 &0 \cr
0 &0 &0
\end{pmatrix}
~~\mbox{and}~~
Y_{u2}=
\begin{pmatrix}0 &0 &0\cr 
*&*&* \cr
*&*&* 
\end{pmatrix}
~,
\label{eq:Yform}
\end{align}
where $*$ denotes a generally non-zero entry.

As in the 2HDM's, one can rotate the Higgs doublet fields into the Higgs basis:
\begin{eqnarray}
\left(\begin{array}{c}
     \Phi_1        \\[1mm]
     \Phi_2         
             \end{array}\right)\,
= R_\beta \left(\begin{array}{c}
     \Phi^{\rm SM}              \\[1mm]
     \Phi'   
             \end{array}\right)\, ,~~ &{\rm with}& ~~ 
R_{\theta}=\left(\begin{array}{cc}
     \cos\theta & -\sin\theta       \\[1mm]
     \sin\theta &  \cos\theta       
             \end{array}\right)
             ~,
\end{eqnarray}
where the new Higgs fields
\begin{eqnarray}
\Phi^{\rm SM} 
=\left(\begin{array}{c}
     G^+       \\[1mm]
     (v+h^{\rm SM}+iG^0)/\sqrt{2}       
             \end{array}\right)\,
~~\mbox{and}~~
\Phi'
=\left(\begin{array}{c}
     H^+       \\[1mm]
     (h'+iA^0)/\sqrt{2}       
             \end{array}\right)
~.   \nonumber
\end{eqnarray}
The mass eigenstates of the CP-even neutral Higgs bosons, $H$ and $h$ with $m_h = 125~{\rm GeV} < m_H$, are
relative to $h^{\rm SM}$ and $h'$ through a rotation:
\begin{eqnarray}
\left(\begin{array}{c}
     H       \\[1mm]
     h         
             \end{array}\right)\,
= R_{\beta-\alpha} \left(\begin{array}{c}
     h^{\rm SM}              \\[1mm]
     h' 
             \end{array}\right)\, .
\end{eqnarray}
In this basis, the original Lagrangian is written as:
\begin{align}
{\cal L}=&
-\Phi^{\rm SM} \bar{u}_{Rj}[Y^{\rm SM}_{u}]_{ji}Q_i
-\Phi' \bar{u}_{Ra}[Y'_{u}]_{ji}Q_i 
-\Phi^{\rm SM} \bar{e}_{Rj}[Y^{\rm SM}_{e}]_{ji}L_i
-\Phi' \bar{e}_{Rj}[Y'_{e}]_{ji}L_i \nonumber \\
&-\Phi^{\rm SM} \bar{d}_{Rj}[Y^{\rm SM}_{d}]_{ji}Q_i
-\Phi' \bar{d}_{Rj}[Y'_{d}]_{ji}Q_i + {\rm h.c.} 
~,
\end{align}
where, $Y^{\rm SM}_{u} = c_\beta Y_{u1} + s_\beta Y_{u2}$.  Throughout this paper, we will use the shorthand notation: $s_\theta = \sin\theta$ and $c_\theta = \cos\theta$.  With the explicit forms in Eq.~\eqref{eq:Yform}, we have
\begin{align}
\begin{split}
Y'_{u} &= -s_\beta Y_{u1} + c_\beta Y_{u2}=
\left(\begin{array}{ccc}
     -\tan\beta  &           &     \\[1mm]
                 & \cot\beta &    \\[1mm]
                 &           & \cot\beta      
             \end{array}\right) Y^{\rm SM}_{u}
~, \\
Y'_e &= -\tan\beta~~ Y^{\rm SM}_e
~, \\
Y'_d &= \cot\beta~~ Y^{\rm SM}_d
~.
\end{split}
\end{align}
The up-type quark mass matrix can be diagonalized via a bi-unitary transformation, $V_u M_u U^{\dagger}={\rm diag}(m_u,m_c,m_t)\equiv v Y^{\rm diag}_u / \sqrt{2}$, with the unitary matrix $V_u$ defined by
\begin{eqnarray}
\label{eqV}
\left(\begin{array}{c}
     u_R             \\[1mm] 
     c_R            \\[1mm]
     t_R                       
             \end{array}\right)_{\rm mass}
=V_u
\left(\begin{array}{c}
     u_{R1}^{Q^{PQ}=1}             \\[1mm] 
     u_{R2}^{Q^{PQ}=0}            \\[1mm]
     u_{R3}^{Q^{PQ}=0}          
             \end{array}\right)
\,.
\end{eqnarray}
In this mass basis,
\begin{align}
\label{eq5}
Y'^{\rm ,diag}_u=
\left(\begin{array}{ccc}
     -\tan\beta  &           &     \\[1mm]
                 & \cot\beta &    \\[1mm]
                 &           & \cot\beta      
             \end{array}\right) Y^{\rm diag}_{u}
-(\tan\beta+\cot\beta)H_uY^{\rm diag}_{u}
~,
\end{align}
where
\begin{eqnarray}
\label{eq6}
H_u=
\left(\begin{array}{ccc}
H^{uu}_u  
& H^{uc}_u 
& H^{ut}_u    \\[1mm]
H^{uc}_u
& H^{cc}_u
& H^{ct}_u   \\[1mm]
 H^{ut}_u
& H^{ct}_u
& H^{tt}_u
             \end{array}\right)
\equiv V_u
\left(\begin{array}{ccc}
     1  &      &     \\[1mm]
        & 0    &    \\[1mm]
        &      & 0    
             \end{array}\right) V_u^{\dagger}
-\left(\begin{array}{ccc}
     1  &      &     \\[1mm]
        & 0    &    \\[1mm]
        &      & 0    
             \end{array}\right)
~,
\end{eqnarray}
and the non-trivial flavor structure among the $u_R, c_R$, and $t_R$ fields is encoded in the matrix $H_u$.  Assuming no additional CP phase for simplicity and without loss of generality, we can parametrize the mixing matrix $V_u$ with a $u-t$ mixing angle $\rho_u$ and a $u-c$ mixing angle $\psi_u$ as
\begin{eqnarray}
\label{eq10}
V_u=
\left(\begin{array}{ccc}
     \cos{\frac{\psi_u}{2}}     & \sin{\frac{\psi_u}{2}}   & 0             \\[1mm] 
     -\sin{\frac{\psi_u}{2}}    & \cos{\frac{\psi_u}{2}}   & 0             \\[1mm]
     0                          &  0                       & 1  
             \end{array}\right)
\times
\left(\begin{array}{ccc}
     \cos{\frac{\rho_u}{2}}  & 0 & \sin{\frac{\rho_u}{2}}    \\[1mm]
   0                       & 1 & 0                       \\[1mm]
   -\sin{\frac{\rho_u}{2}}   & 0 & \cos{\frac{\rho_u}{2}}    
             \end{array}\right)
~.
\end{eqnarray}
The first and second matrices on the right-hand side of the above equation represent respectively the mixing between $u_R$ and $c_R$ and the mixing between $u_R$ and $t_R$.
In the next section, we will see that both $\rho_u$ and $\psi_u$ are constrained to be close to $0$ or $\pi$.
We note that $(\rho_u,\psi_u) \approx (0,0)$ corresponds to a PQ-charged up quark, $(\rho_u,\psi_u) \approx (0,\pi)$ to a PQ-charged charm quark, and $(\rho_u,\psi_u) \approx (\pi, 0$ or $\pi)$ to a PQ-charged top quark.  The last case is identical to the scenario discussed in Ref.~\cite{Chiang:2017fjr}, where $\tan\beta$ is restricted to moderate values and the muon $g-2$ cannot be explained.  With this parameterization, the explicit form of $H_u$ becomes
\begin{eqnarray}
H_u=
\left(\begin{array}{ccc}
\frac{1+c_{\psi_u}}{2}\frac{1+c_{\rho_u}}{2}-1 ~~ 
& \frac{-s_{\psi_u}}{2} \frac{1+c_{\rho_u}}{2}
& \frac{-c_{\psi_u/2}s_{\rho_u}}{2}    \\[1mm]
\frac{-s_{\psi_u}}{2} \frac{1+c_{\rho_u}}{2}
& \frac{1-c_{\psi_u}}{2}\frac{1+c_{\rho_u}}{2}
& \frac{s_{\psi_u/2}s_{\rho_u}}{2}   \\[1mm]
 \frac{-c_{\psi_u/2}s_{\rho_u}}{2}
& \frac{s_{\psi_u/2}s_{\rho_u}}{2}
& \frac{1-c_{\rho_u}}{2}
             \end{array}\right).
\end{eqnarray}

The Yukawa Lagrangian can now be cast into a simpler form:
\begin{equation}
\label{eq11}
\mathcal{L}\supset \sum_{f,f'}^{u,c,t,d,s,b,e,\mu,\tau}-\frac{m_{f'}}{v}
(\xi^h_{ff'}h\bar{f}_Rf'_L+\xi^H_{ff'}H\bar{f}_Rf'_L
+i\xi^{A}_{ff'}A^0\bar{f}_Rf'_L)+{\rm h.c}\,,
\end{equation}
where
\begin{align}
\begin{split}
\xi^h_{ff'} &\equiv s_{\beta-\alpha}\delta_{ff'} +c_{\beta-\alpha}\zeta_{ff'}\,, 
\\
\xi^H_{ff'} &\equiv c_{\beta-\alpha}\delta_{ff'} -s_{\beta-\alpha}\zeta_{ff'}\,,  
\\
\xi^A_{ff'} &\equiv (2T^f_3) \zeta_{ff'} ~,
\end{split}
\end{align}
and
\begin{eqnarray}
\zeta_{ff'} =
\begin{cases}  
-\tan\beta \, \delta_{ff'}- (\tan\beta+\cot\beta) H_{u, ff'} \,~~~~({\rm for}~f=u) ~, \\
\ \ \ \cot \beta \, \delta_{ff'}- (\tan\beta+\cot\beta) H_{u, ff'} \,~~~~({\rm for}~f=c,t) ~,\\
\ \ \ \cot\beta \, \delta_{ff'}\,~~~~~~~~~~~~~~~~~~~~~~~~~~~~~~~~~~~({\rm for}~f=d,s,b) ~, \\
-\tan\beta \, \delta_{ff'}\,~~~~~~~~~~~~~~~~~~~~~~~~~~~~~~~~~~~({\rm for}~f=e,\mu,\tau) ~,
\end{cases}
\end{eqnarray}
where $T^3_f$ denotes the $SU(2)_L$ eigenvalue of a fermion $f$ ({\it i.e.}, $T^3_f=+1/2$ for $u,c,t$, and $-1/2$ for $d,s,b,e,\mu,\tau$).
The FCNS interactions only appear in the up-quark sector, and the these interactions can be written as
\begin{eqnarray}
\label{eq13}
\mathcal{L} \supset  \sum^{f\neq f'}_{f,f'=u,c,t} \frac{m_{f'}}{v}
(\tan\beta+\cot\beta)H_{u,ff'}
\left[
c_{\beta-\alpha}
h\bar{f_R}f'_L
-s_{\beta-\alpha}
H\bar{f_R}f'_L
+iA^0\bar{f_R}f'_L  
\right]
+{\rm h.c.}
\end{eqnarray}
The pattern of the FCNS interactions of $h, H$ is hence identical to that of $A$ but suppressed 
by $c_{\beta-\alpha}$ and $s_{\beta-\alpha}$, respectively. 
Although we parametrize the mixing matrix $V_u$ and $H_u$ using two mixing angles $\rho_u$ and $\psi_u$, we will see in the next section that the $u-c$ mixing angle $\psi_u$ receives stronger constraints than the $u-t$ mixing angle $\rho_u$ from the $D-\bar{D}$ oscillation. Therefore, it is reasonable to fix $\psi_u = 0$ in the mixing matrix $V_u$
for other phenomenological studies. 
In this limit, we have 
\begin{align}
\begin{split}
\zeta_{uu} &\equiv -\tan\beta - (\tan\beta+\cot\beta) \frac{\cos\rho_u -1}{2} ~, \\
\zeta_{cc} &\equiv \cot\beta ~, \\
\zeta_{tt} &\equiv \cot\beta - (\tan\beta+\cot\beta)\frac{1-\cos\rho_u }{2} ~,\\
\zeta_{ut} &= \zeta_{tu} = (\tan\beta+\cot\beta)\frac{\sin\rho_u}{2} ~.
\end{split}
\end{align}

\section{Constraints}
\label{sec:current_bounds}

In this section we consider the current experimental constraints on this model.  We examine first the constraints from low energy observables: muon $g-2$, lepton universality in $\tau$ and $Z$ decays, and the $B_s \to \mu\mu$ decay.  We perform a $\chi^2$-fit to those four observables to find the preferred parameter region. We consider the observables one by one as follows.

\subsection*{Muon $g-2$} 

The discrepancy between the experimental measurement~\cite{Bennett:2006fi} 
and the SM prediction of the muon anomalous magnetic moment~\cite{Jegerlehner:2009ry} is a long-standing puzzle, 
and the deviation~\cite{Broggio:2014mna}
\begin{equation}
\Delta a_\mu = a^{\rm EXP}_\mu-a^{\rm SM}_\mu=(262\pm 85) \times 10^{-11}
\end{equation}
is at about $3.1\sigma$ level.
Additional 1-loop contributions from the $h,H,A,H^{\pm}$ bosons in the VAM are the same as those in the usual 2HDM~\cite{2HDMg2}, and the sum is given by
\begin{equation}
\Delta a^{\rm VAM,1-loop}_\mu
=\frac{G_F m^2_\mu}{4 \sqrt{2} \pi^2 }
\sum_{i}^{h,H,A,H^{\pm}} (\xi^i_{\mu\mu})^2r^i_\mu f_i(r^i_\mu)\,,
\label{eq:1loopVAM}
\end{equation} 
where $G_F$ is the Fermi decay constant, $m_\mu$ is the muon mass, $\alpha_{\rm em}$ is the fine structure constant, $r^j_f=m^2_f/m^2_j$ and the loop functions
\begin{eqnarray}
f_{h,H}(r)\! =\!\!\int^{1}_0\!\!\! dx \frac{x^2(2-x)}{1-x+rx^2} ~,\ \ 
f_{A}(r)\! =\!\!\int^{1}_0\!\!\!  dx \frac{-x^3}{1-x+rx^2} ~, \ \ 
f_{H^{\pm}}(r)\! =\!\!\int^{1}_0\!\!\!  dx \frac{-x(1-x)}{1-r(1-x)} ~. \nonumber
\end{eqnarray}
Note that the sign of each contribution in Eq.~\eqref{eq:1loopVAM} is solely determined by that of the corresponding loop function.   In particular, the diagram associated with $A$ ($H$) gives the leading negative (positive) contribution among all.

The 2-loop Barr-Zee contributions may also be important in this model.  Those involving the heavy fermions have the contribution
\begin{equation}
\Delta a^{\rm VAM, BZ}_\mu
=\frac{G_F m^2_\mu}{4 \sqrt{2} \pi^2}\frac{\alpha_{\rm em}}{\pi}
\sum_{i}^{h,H,A}\sum_{f}^{t,b,c,\tau}N^c_fQ^2_f \xi^i_{\mu\mu}\xi^i_{ff}
r^i_fg_i(r^i_f)\,,
\end{equation}
where the loop functions are defined as 
\begin{eqnarray}
g_{h,H}(r) \!= \!\! \int^1_0 \!\!\!dx \frac{2x(1-x)-1}{x(1-x)-r}{\rm ln} \frac{x(1-x)}{r}
~,~~  
g_A(r) \!= \!\! \int^1_0  \!\!\!dx \frac{1}{x(1-x)-r}{\rm ln} \frac{x(1-x)}{r}
~. \nonumber
\end{eqnarray}

The explicit values of the 1-loop contribution and the 2-loop Barr-Zee contributions for the case of $m_H=m_A=1~\tev$ are shown in Table~\ref{tab:g-2contribution}.  To obtain $\Delta a_\mu$ from the table, one needs to multiply a common pre-factor of $G_Fm_\mu^2 / (4\sqrt{2}\pi^2) =  2.3 \times 10^{-9}$. 
To ameliorate the $3.1\sigma$ deviation, the total contribution from the 
VAM $\Delta a^{\rm VAM}_\mu = \Delta a^{\rm VAM,1-loop}_\mu + \Delta a^{\rm VAM,BZ}_\mu$ 
must be about ${\cal O}(10^{-9})$ and positive.
Therefore, we require an ${\cal O}(1)$ positive contribution without the pre-factor.
The loop functions are monotonically increasing function of $m_\phi$ while the mass dependence is not strong as long as $r^\phi_f \ll 1$. 
Hence, it is a good approximation that $\Delta a_\mu \propto m_{\phi}^{-2}$ both for 1-loop and 2-loop contributions.
In Type-X 2HDM, for example, taking $m_A \sim 30~\gev$ and $\tan \beta \sim 40$ renders a factor of $10^6$ enhancement on the 2-loop $\tau$ contribution, which leads to the required size of ${\cal O}(1)$ positive contribution.

\begin{table}[h!]
\begin{tabular}{lc|c|c|c|c}
 & \ fermion \  &  $(g^H_f, g^A_f)$ & $(r^H_f g^H_f, r^A_f g^A_f)$ &$\times \alpha N_f^c Q_f^2/\pi$& {\rm sign} of $(\delta_H, \delta_A)$\\
  \hline\hline
1-loop & $\mu $ & $(17, -16)$ & $(1.9, -1.8) \cdot 10^{-7}$   & $(1.9, -1.8) \cdot 10^{-7}$ & $(+, -)$  \\  
 \hline
&  $t$  & $(-12,15.9)$& $(-3.6, 4.7) \cdot 10^{-1}$   & $(-1.1, 1.5) \cdot 10^{-3}$ & $(-, -)$  \\
& $c$ & $(-118,140)$ & $(-1.9, 2.3) \cdot  10^{-4}$  & $(-5.9,7.1) \cdot 10^{-7}$ & $(-, -)$ \\
2-loop & $u$ & $(-282,330)$  & $(-1.5, 1.7) \cdot  10^{-9}$  & $(-4.6,5.4) \cdot 10^{-12}$ & $(-, -)$ \\
& $b$ & $(-87, 105)$ &$(-1.5, 1.8) \cdot 10^{-3}$ & $(-1.1,1.4) \cdot 10^{-6}$ & $(-, +)$  \\
& $\tau$ & $(-109, 130)$  & $(-3.4, 4.1) \cdot  10^{-4}$  & $(-8.0,9.6) \cdot 10^{-7}$ & $(-, +)$ \\
 \hline
 \end{tabular}
\label{tab:g-2contribution}
\caption{Explicit values of the selected loop functions for $m_H = m_A = 1~\tev$.  The numbers in the first column show $g^\phi_f \equiv f_\phi(r^\phi_f)$ for 1-loop and $g^\phi_f \equiv g_\phi(r^\phi_f)$ for 2-loop cases ($\phi=H,A$).  The sign of the contribution to $(g-2)_\mu$ is shown, including the $T_3^\mu T_3^f$ factor, for the $A$ contributions.
Note that only the $t, c, u$ contributions for $A$ change the sign while those of $b, \tau$ do not.}
\end{table}

The sign of each contribution is determined by the corresponding $\xi^\phi_{\mu\mu}\xi^\phi_{ff}$, which is proportional to $\zeta_{\mu\mu}\zeta_{ff}$ for the $A$ contributions and for $H$ in the aligned limit ($c_{\beta-\alpha}=0$). 
The last column in the table summarizes the sign of each contribution $\delta_{H/A}$ modulo 
$\zeta_{\mu\mu}\zeta_{ff}$.  
For the parameter region of $m_A< m_H=m_{H}^\pm$, the 1-loop contribution is always negative.  Hence, a significant positive contribution is required from the 2-loop Barr-Zee diagrams.  The Barr-Zee diagram contribution involving a light $A$ is proportional to $\zeta_{ff}=-\tan\beta$ or $\cot\beta$.  With $\zeta_{\mu\mu}=- \tan\beta$, we see from the table that only the $\tau$-loop offers a positive contribution among the $\tan^2\beta$ enhanced contributions. 
With a large $\tan\beta$ enhancement, it dominates 
to compensate for the other negative 1-loop and 2-loop contributions.
In the up-type VAM, the bottom loop contribution is negative but negligible since $\tan\beta$ is canceled by $\cot\beta$.
We find that the charm-specific VAM is not preferred as the charm loop contribution is negative and enhanced by $\tan^2\beta$, 
while the up-specific VAM is still viable as one can neglect the small up Yukawa coupling.

\begin{figure}[h!tb]
\centering
\includegraphics[height=1.5in,angle=0]{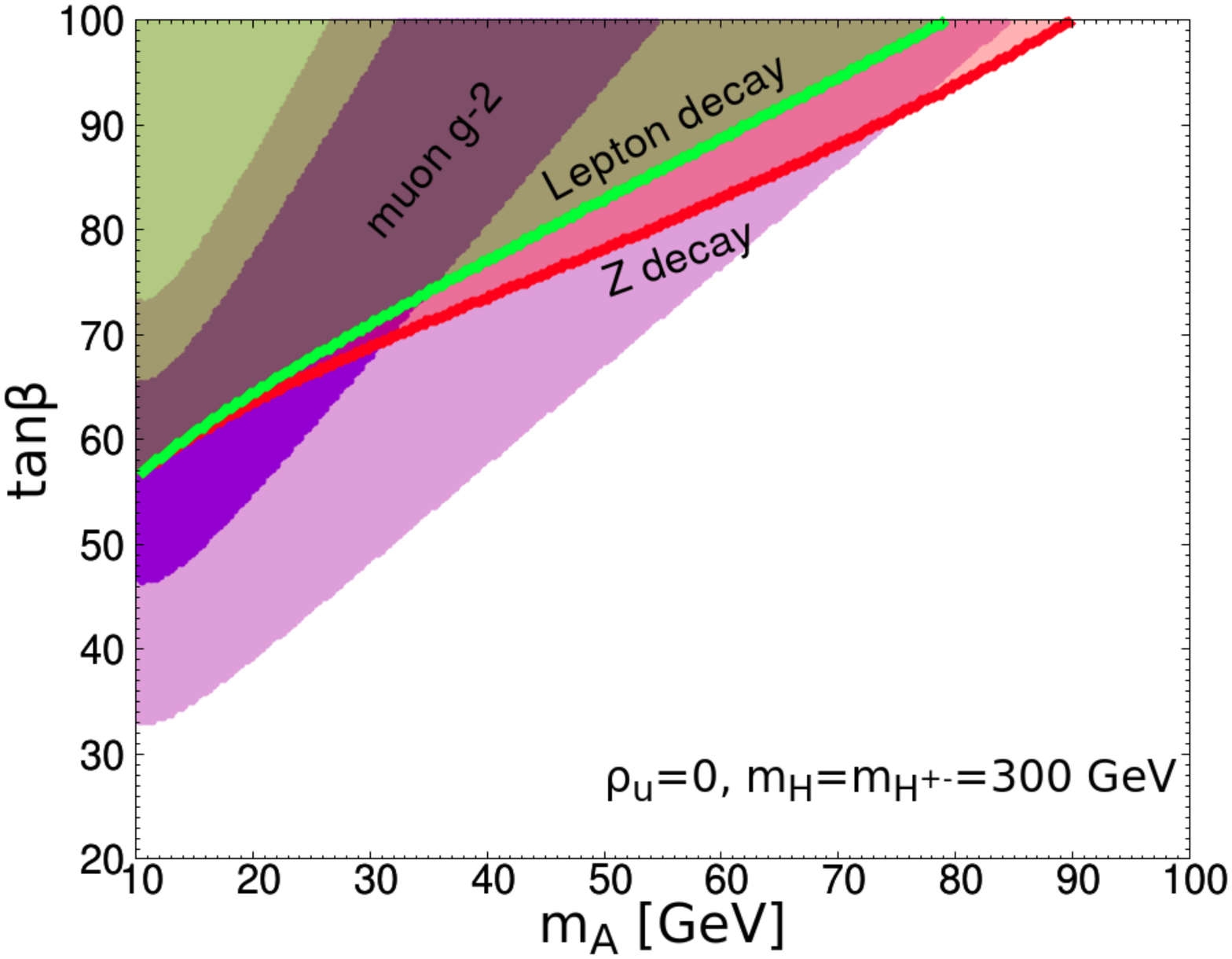}
\includegraphics[height=1.5in,angle=0]{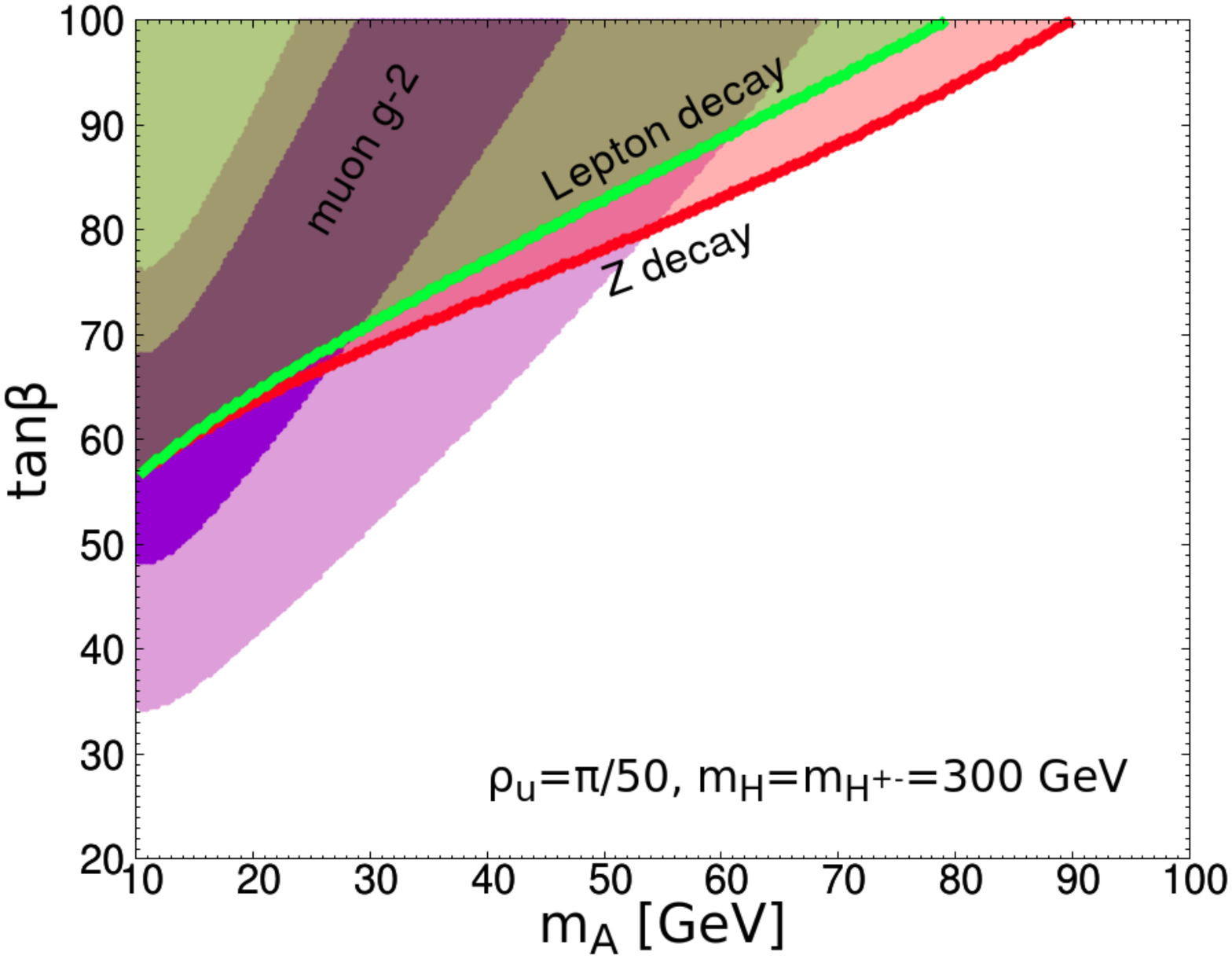}
\includegraphics[height=1.5in,angle=0]{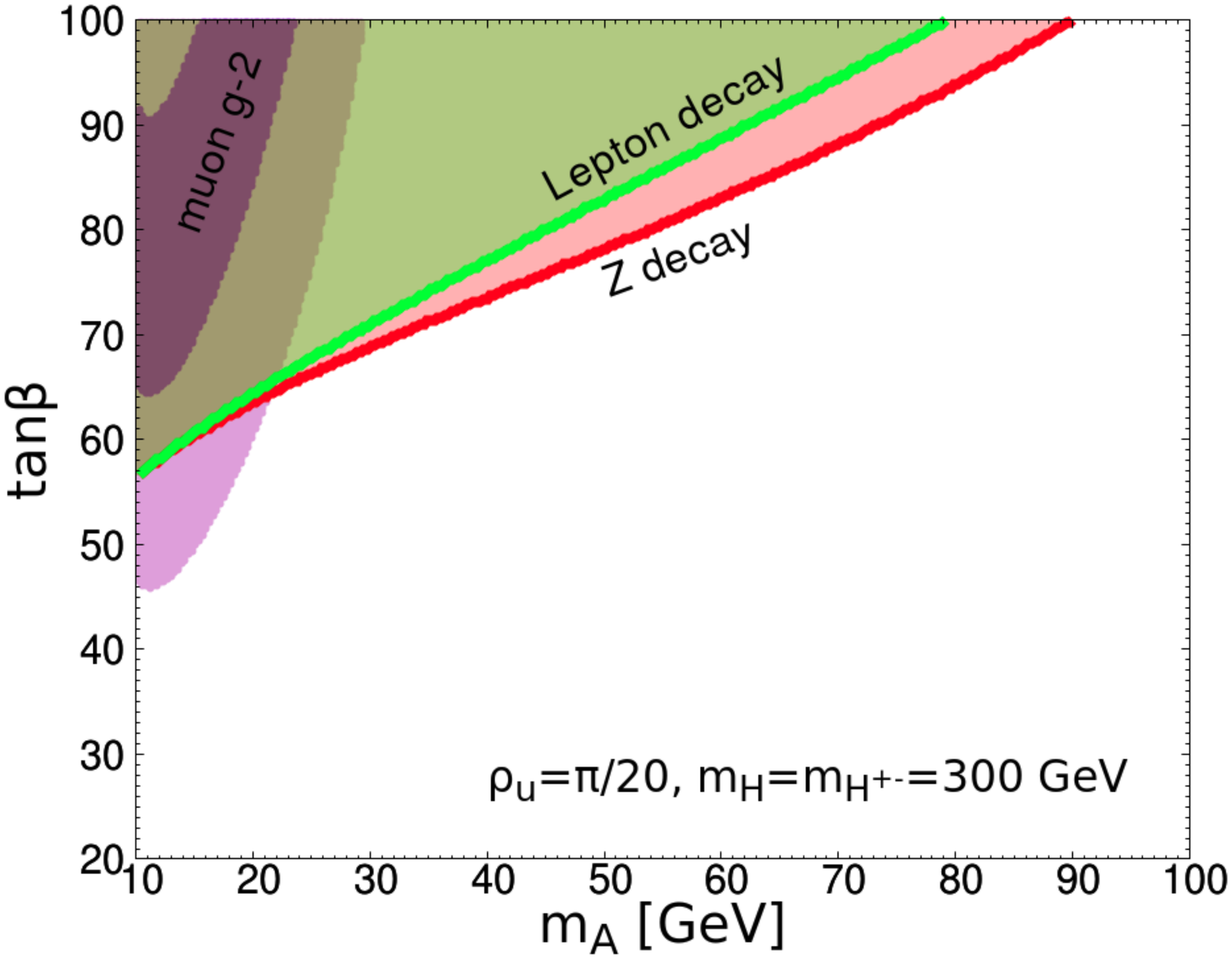}
\caption{\small \label{fig:g-2} 
Allowed parameter region in the $m_A$-$\tan\beta$ plane for $\rho_u=0$ (left), $\frac{\pi}{50}$ (middle), and $\frac{\pi}{20}$ (right).
The $1\sigma$ and $2\sigma$ regions preferred by the muon $g-2$ is drawn in dark and light purple, respectively~\cite{2HDMg2}. 
The constraint of lepton universality in the $\tau$~\cite{2HDMg2} ($Z$~\cite{zdecay}) decays is given by the green (red) curve, and the region above it is excluded at 95\% confidence level (CL). 
}
\end{figure}

Moreover, any non-zero mixing between $u$ and $t$, $\rho_u \neq 0$, does not help explaining the muon $g-2$ 
in the up-specific VAM because a non-zero $\rho_u$ always reduces $\xi^A_{tt}$ and may turn the originally positive top-loop Barr-Zee contribution negative.   Therefore, a large value of $\rho_u$ is not favored. 
This is also seen in the charm-specific VAM for the same reason.  
The parameter region consistent with the muon $g-2$ observation is shown in purple in Fig.~\ref{fig:g-2} for $\rho_u = 0, \pi/50$, and $\pi/20$.

\subsection*{Lepton universality in $\tau,\mu$ decays} 

The lepton decay processes $\tau \to \mu\nu\nu, \tau \to e\nu\nu$, 
and $\mu \to e\nu\nu$ can be used to constrain this model, as they can be mediated at tree level by $H^\pm$ in addition to the $W^\pm$ bosons.  This is true for any 2HDM's in general.  There are also loop contributions mediated by $h,H,A$ and $H^\pm$~\cite{2HDMg2}.
The Heavy Flavor Averaging Group (HFAG) gives constraints on the coupling ratios
$g_{\ell}/g_{\ell'}$, where $\ell,\ell'=e, \mu,\tau$~\cite{hfag}.  
Here we quote the deviations of these ratios from their SM values, $\bar{\delta}_{\ell\ell'}\equiv (g_\ell/g_{\ell'})-1$:
$\bar{\delta}_{\tau\mu}=0.0011\pm0.0015$, 
$\bar{\delta}_{\tau e}=0.0029\pm0.0015$, and
$\bar{\delta}_{\mu e}=0.0018\pm0.0014$.

By neglecting the electron mass, we obtain 
$\bar{\delta}_{\tau\mu}=\delta^{\rm VAM}_{\rm loop}$, 
$\bar{\delta}_{\tau e}=\delta^{\rm VAM}_{\rm tree}+\delta^{\rm VAM}_{\rm loop}$, 
and $\bar{\delta}_{\mu e}=\delta^{\rm VAM}_{\rm tree}$, with
\begin{eqnarray}
\delta^{\rm VAM}_{\rm tree}&=& \frac{m^2_\tau m^2_\mu}{8m^4_{H^\pm}}t^4_\beta
-\frac{m^2_\mu}{m^2_{H^\pm}}t^2_\beta
\frac{g(m^2_\mu/m^2_\tau)}{f(m^2_\mu/m^2_\tau)} \,, \nonumber \\
\delta^{\rm VAM}_{\rm loop}&=&\frac{G_Fm^2_\tau}{8\sqrt{2}\pi^2}t^2_\beta
\left[1
+\frac{1}{4}
\left( H(x_A)+s^2_{\beta-\alpha}H(x_H)+c^2_{\beta-\alpha}H(x_h) \right) \right]\,, 
\end{eqnarray}
where the loop functions 
$f(x)=1-8x+8x^3-x^4-12x^2{\rm ln}(x)
,g(x)=1+9x-9x^2-x^3+6x(1+x){\rm ln(x)}
,H(x)={\rm ln}(x)(1+x)/(1-x)$, and $x_{A,H,h}=m^2_{A,H,h}/m^2_{H^\pm}$.
Taking correlations among the observables into account, we use the following three independent quantities 
for the $\chi^2$-fit~\cite{2HDMg2}
\begin{align}
\begin{split}
\sqrt{\frac{3}{2}}\delta^{\rm VAM}_{\rm tree} &= 0.0022\pm 0.0017
~, \\  
\delta^{\rm VAM}_{\rm loop} &= 0.0001\pm 0.0014
~, \\
\frac{1}{\sqrt{2}}\delta^{\rm VAM}_{\rm tree}+\sqrt{2}\delta^{\rm VAM}_{\rm loop}
&= 0.0028\pm 0.0019
~.
\end{split}
\end{align}

Both $\delta_{\rm tree}^{\rm VAM}$ and $\delta_{\rm loop}^{\rm VAM}$ are negative in the VAM, while the observed data prefer to have positive values.  Thus, we can set an upper bound on $\tan\beta$ for a fixed value of $m_A$.  The 95\% CL excluded region is overlaid as the green area in Fig.~\ref{fig:g-2}.   
Note that this constraint is independent of the quark sector and thus the quark mixing parameter $\rho_u$.

\subsection*{Lepton universality in $Z$ decays}

The precision measurements at the $Z$-pole in both SLD and LEP experiments provide ratios of the leptonic $Z$ decay branching fractions~\cite{ALEPH:2005ab}.
Consider the deviations of such ratios from identity, defined by $\delta_{\ell\ell}\equiv (\Gamma_{Z\to \ell^+\ell^-}/\Gamma_{Z\to e^+e^-})-1$.  Current data have
\begin{eqnarray}
\delta_{\mu\mu}=0.0009\pm0.0028 \ \ \ \  {\rm and}\ \ \ \  \delta_{\tau\tau} = 0.0019\pm0.0032
~.
\end{eqnarray}
Corrections due to the $A,H,H^{\pm}$ loops in the VAM are found to be~\cite{zdecay}
\begin{eqnarray}
\delta^{\rm VAM}_{\mu\mu} \simeq 0 ~, \ \ \ \  {\rm and}\ \ \ \ 
\delta^{\rm VAM}_{\tau\tau} = \frac{2g^e_L{\rm Re}(\delta g^{\rm VAM}_L)+2g^e_R{\rm Re}(\delta g^{\rm VAM}_L)}{(g^e_L)^2+(g^e_R)^2}\,,
\end{eqnarray}
where the SM couplings $g^e_L=-0.27$ and $g^e_R=0.23$.
The corrections $\delta g^{\rm VAM}_L$ and $\delta g^{\rm VAM}_L$
from the VAM's are given by
\begin{eqnarray}
\delta g^{\rm VAM}_L &=& \frac{1}{16\pi^2}
\left(\frac{m_\tau}{v}\xi^A_{\tau\tau}\right)^2
\left\lbrace
-\frac{1}{2}B_Z(r_A)-\frac{1}{2}B_Z(r_H)-2C_Z(r_A,r_H)
\right. \nonumber \\
&&\left.
+s^2_W\left[B_Z(r_A)+B_Z(r_H)+\tilde{C}_Z(r_A)+\tilde{C}_Z(r_H) \right]
\right\rbrace \,,\nonumber \\
\delta g^{\rm VAM}_R &=&\frac{1}{16\pi^2}
\left(\frac{m_\tau}{v}\xi^A_{\tau\tau}\right)^2
\left\lbrace
2C_Z(r_A,r_H)-2C_Z(r_{H^\pm},r_{H^\pm})+\tilde{C}_Z(r_{H^\pm})
-\frac{1}{2}\tilde{C}_Z(r_A)-\frac{1}{2}\tilde{C}_Z(r_H)
\right. \nonumber \\
&&\left.
+s^2_W\left[ B_Z(r_A)+B_Z(r_H)+2B_Z(r_{H^\pm})+\tilde{C}_Z(r_A)
+\tilde{C}_Z(r_H)+4C_Z(r_{H^\pm},r_{H^\pm}) \right]
\right\rbrace \,, \nonumber
\end{eqnarray}
where $r_{A,H,H^\pm}=m^2_{A,H,H^\pm}/m^2_Z$, $s^2_W \equiv \sin^2\theta_W \simeq 0.23$, and the loop functions 
\begin{eqnarray}
B_Z(r)&=& -\frac{\Delta_\epsilon}{2}-\frac{1}{4}+\frac{1}{2} \ln(r) \,, \nonumber \\
C_Z(r_1,r_2) &=& \frac{\Delta_\epsilon}{4}-\frac{1}{2}\int^1_0 dx \int^x_0
dy \ln[r_2(1-x)+(r_1-1)y+xy]\,, \nonumber \\
\tilde{C_Z}(r)&=& \frac{\Delta_\epsilon}{2}+\frac{1}{2}-r[1+\ln(r)]
+r^2[\ln(r)\ln(1+r^{-1})-{\rm dilog}(-r^{-1})] \nonumber \\
&& -\frac{i\pi}{2}[1-2r+2r^2\ln(1+r^{-1})]\,, \nonumber \\
{\rm dilog}(z) &\equiv & \int^1_0 dx \frac{\ln(1-x)}{x}\,.
\end{eqnarray}
The renormalization constant $\Delta_\epsilon=2/\epsilon-\gamma+\ln(4\pi)$ from dimensional regularization will cancel in $\delta g^{\rm VAM}_L$ and $\delta g^{\rm VAM}_R$.

While the contributions from the VAM's are negative, the present data exhibit slightly larger values than the SM predictions.
Therefore, the large $\tan\beta$ region with an enhancement in the $A\tau\tau$ coupling is disfavored.
The region excluded at 95\% CL is overlaid in the red region in Fig.~\ref{fig:g-2}.  
Note that this constraint is insensitive to the quark sector.
From the above considerations, it is seen that small $m_A$ and large $\tan\beta$ are preferred in the up-specific VAM.

\subsection*{Bottom rare decay $B^0_s \to \mu^+\mu^-$}

In the SM, the $b \to s$ flavor-changing neutral current (FCNC) processes are mediated by the loop diagrams with the $W^{\pm}$ boson and top quark.  In the VAM with non-zero quark mixing angle $\rho_u$, the pseudoscalar $A$ can couple to the top quark through the $\xi^A_{tt}$ coupling.
Since $A$ is preferred to be light according to the discussions in the previous 
subsection, its contribution to $B_s \to \mu^+\mu^-$ cannot be neglected.  
The time-integrated branching ratio $\overline{\rm BR}(B^0_s \to \mu^+\mu^-)_{\rm EXP}$ 
averaged between the LHCb~\cite{bs-LHCb} and CMS~\cite{bs-CMS} Collaborations normalized to the SM prediction is reported to be:
\begin{equation}
\bar{R}_{s\mu}\equiv \frac{\overline{\rm BR}(B^0_s \to \mu^+\mu^-)_{\rm EXP}}{\overline{\rm BR}(B^0_s \to \mu^+\mu^-)_{\rm SM}}=0.79 \pm 0.20 ~.
\end{equation}
The combined SM and VAM contribution to this observable is~\cite{bs}
\begin{equation}
\bar{R}_{s\mu}=\left[|P|^2+\left(1-\frac{\Delta \Gamma_s}{\Gamma^s_L} \right)|S|^2 \right]\,,
\end{equation}
where $\Delta \Gamma_s=0.081\,{\rm ps^{-1}}$ and $\Gamma^s_L=1/1.428\,{\rm ps^{-1}}$
are the decay width difference between the two $B_s$ mass eigenstates and the width of the lighter mass eigenstate, respectively.
The pseudoscalar and scalar contributions are given by
\begin{align}
\begin{split}
P &\equiv \frac{C_{10}}{C^{\rm SM}_{10}}
+\frac{M^2_{B_s}}{2M^2_W}\left(\frac{m_b}{m_b+m_s}\right)\frac{C_P-C^{\rm SM}_P}{C^{\rm SM}_{10}}\,,  
\\
S &\equiv \sqrt{1-\frac{4m^2_\mu}{M^2_{B_s}}}
\frac{M^2_{B_s}}{2M^2_W}\left(\frac{m_b}{m_b+m_s}\right)\frac{C_S-C^{\rm SM}_S}{C^{\rm SM}_{10}}\,,
\end{split}
\end{align}
where $C_{10}$, $C_S$, and $C_P$ are the Wilson coefficients of the effective four-fermion operators $\mathcal{O}_{10}$, $\mathcal{O}_{S}$, and $\mathcal{O}_{P}$ that include contributions from both SM and VAM, with the detailed expressions given in Ref.~\cite{bs}.  The measured value of $\bar{R}_{s\mu}$ is slightly smaller at the $1\sigma$ level than the SM prediction $\bar{R}_{s\mu}=1$ out of $P=1$ and $S=0$.

In the up-specific VAM, the main contribution appears in $P$ and one can neglect $S$. 
The contribution from the top-$W$ loop diagrams with a pseudoscalar $A$ propagator is proportional to $-\xi^A_{tt}\xi^A_{\mu\mu}$.
For $\rho_u=0$, the contribution is positive and independent of $\tan\beta$, with $\Delta P \sim 0.21$ for $m_A=15~\gev$. 
As $\rho_u$ increases, it decreases to zero at $\rho_u \simeq 2/\tan\beta$ and eventually becomes negative.
The bottom quark mediated diagrams with a pseudoscalar $A$ propagator contribute negatively as $\Delta P \sim -0.17$ for $m_A=15~\gev$, which is independent of $\tan\beta$ as it is proportional to $-\xi^A_{bb}\xi^A_{\mu\mu}$.
In summary, a small but non-zero $\rho_u$ is preferred to fit the current $B^0_s \to \mu^+\mu^-$ data, which exhibits a small downward deviation $\Delta P \sim -0.1$. There exists another solution corresponding to $\Delta P \sim -1.9$.

\begin{figure}[h]
\centering
\includegraphics[height=2.2in,angle=0]{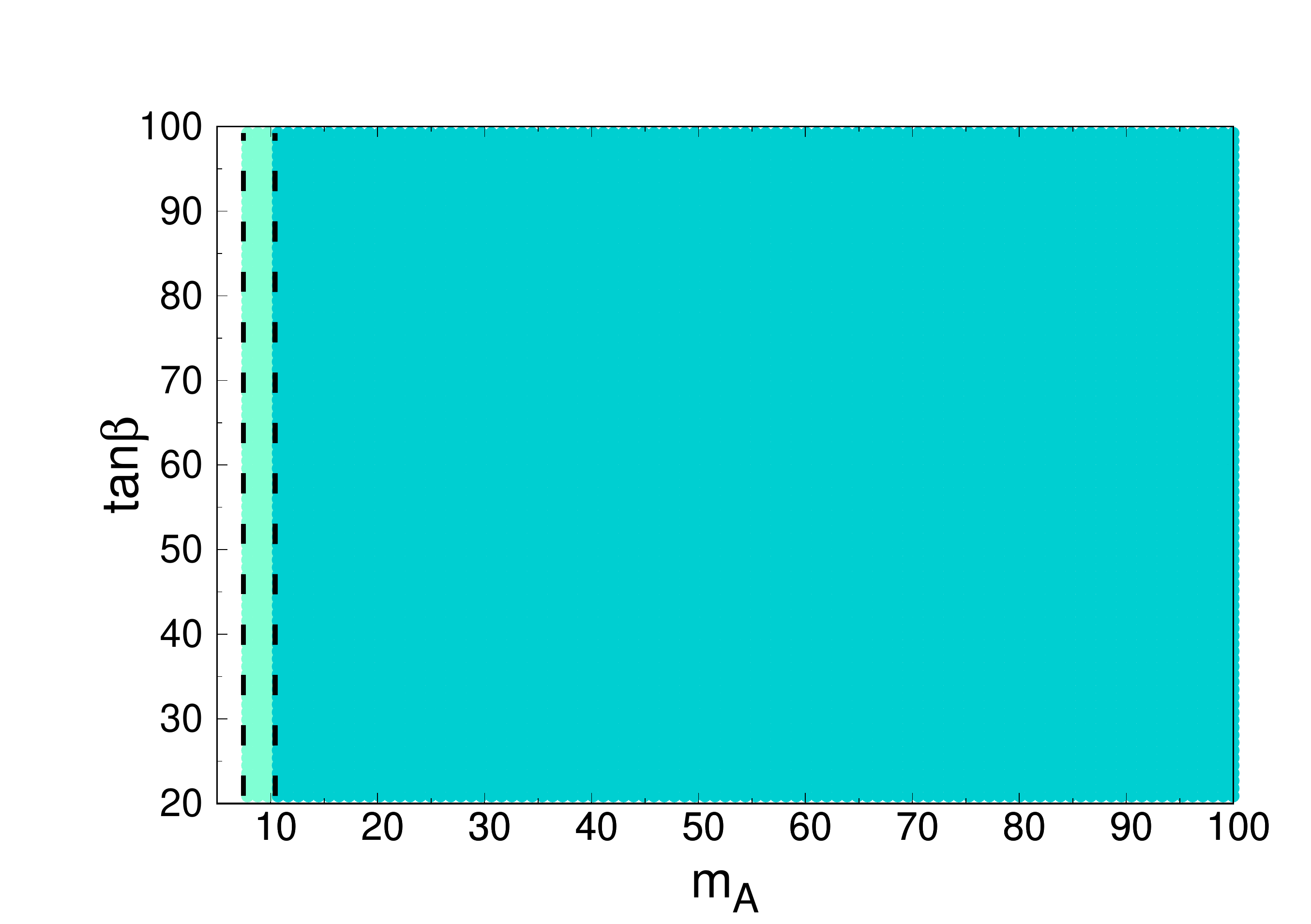}
\includegraphics[height=2.2in,angle=0]{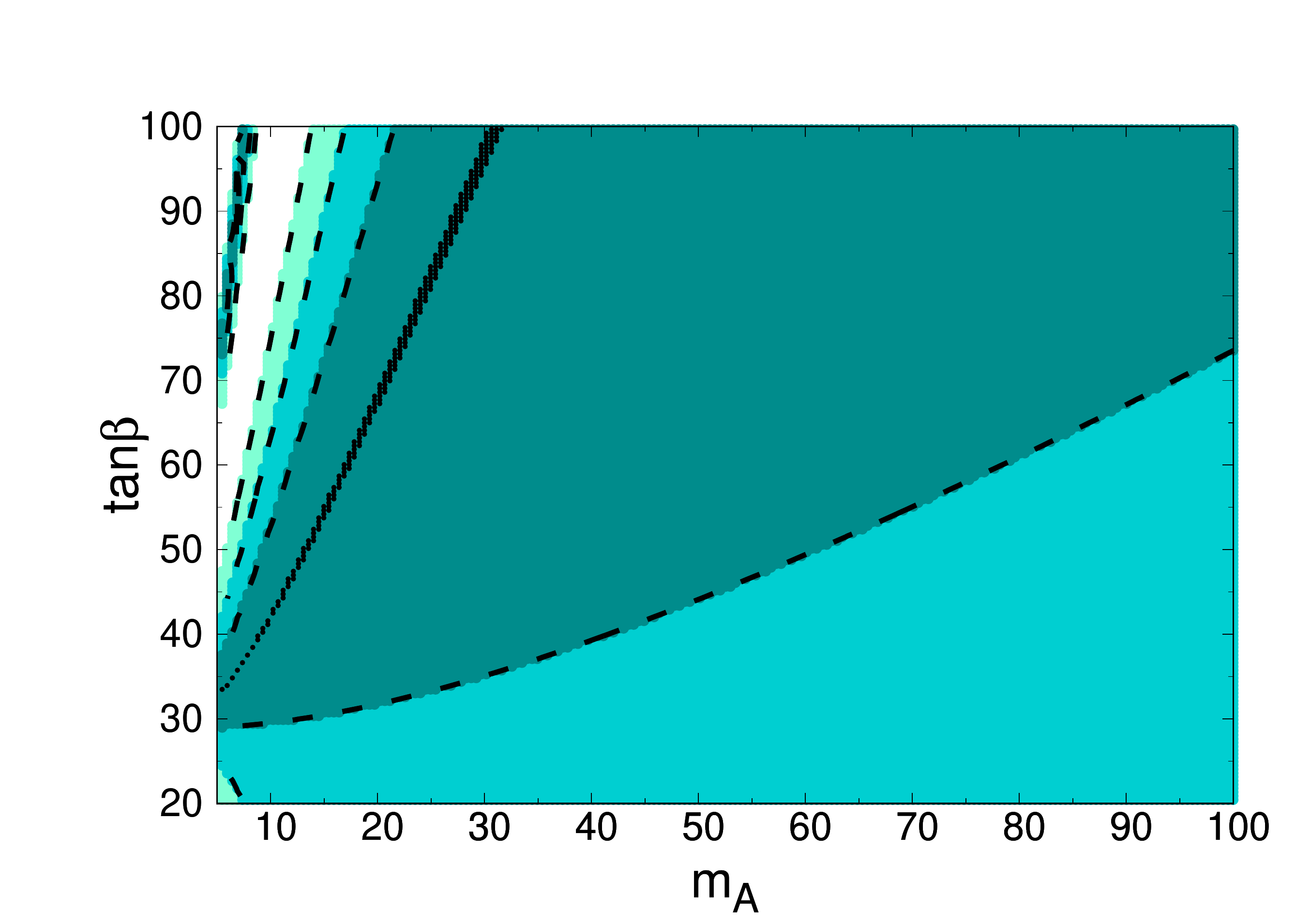}
\vspace{-0.5cm}
\caption{\small \label{fig:chi2Bsmumu} 
Contours of the $\chi^2$-fit value for $B_s\to \mu^+\mu^-$ for $\rho_u = 0$ (left) and $\rho_u = \pi/100$ (right). 
The best-fit curve, $1\sigma$, $2\sigma$ and $3\sigma$ regions are shown in different colors. 
As a comparison, $\chi^2_{\rm SM}=1.1$.
}
\end{figure}

\begin{figure}[h]
\centering
\includegraphics[height=2.2in,angle=0]{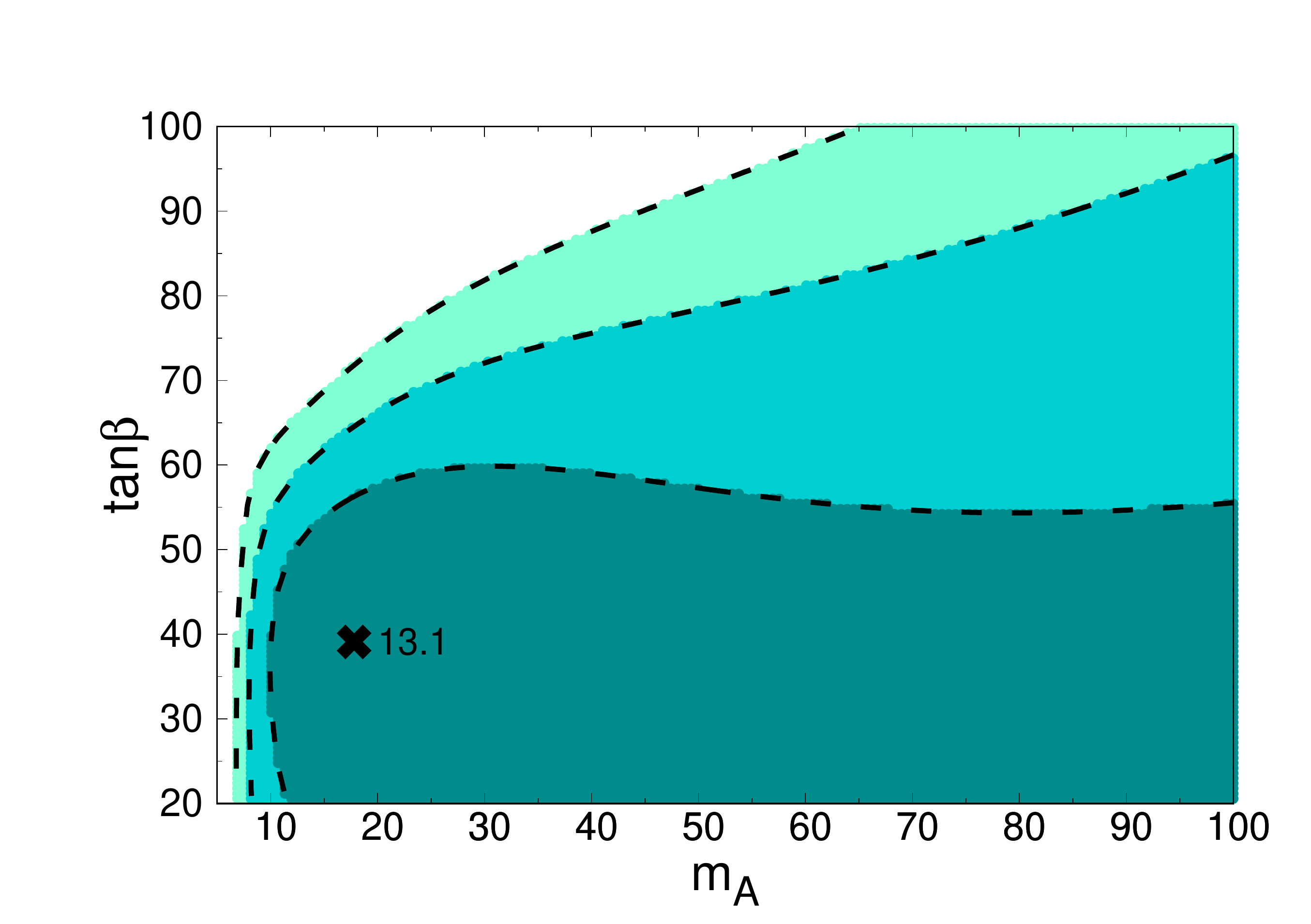}
\includegraphics[height=2.2in,angle=0]{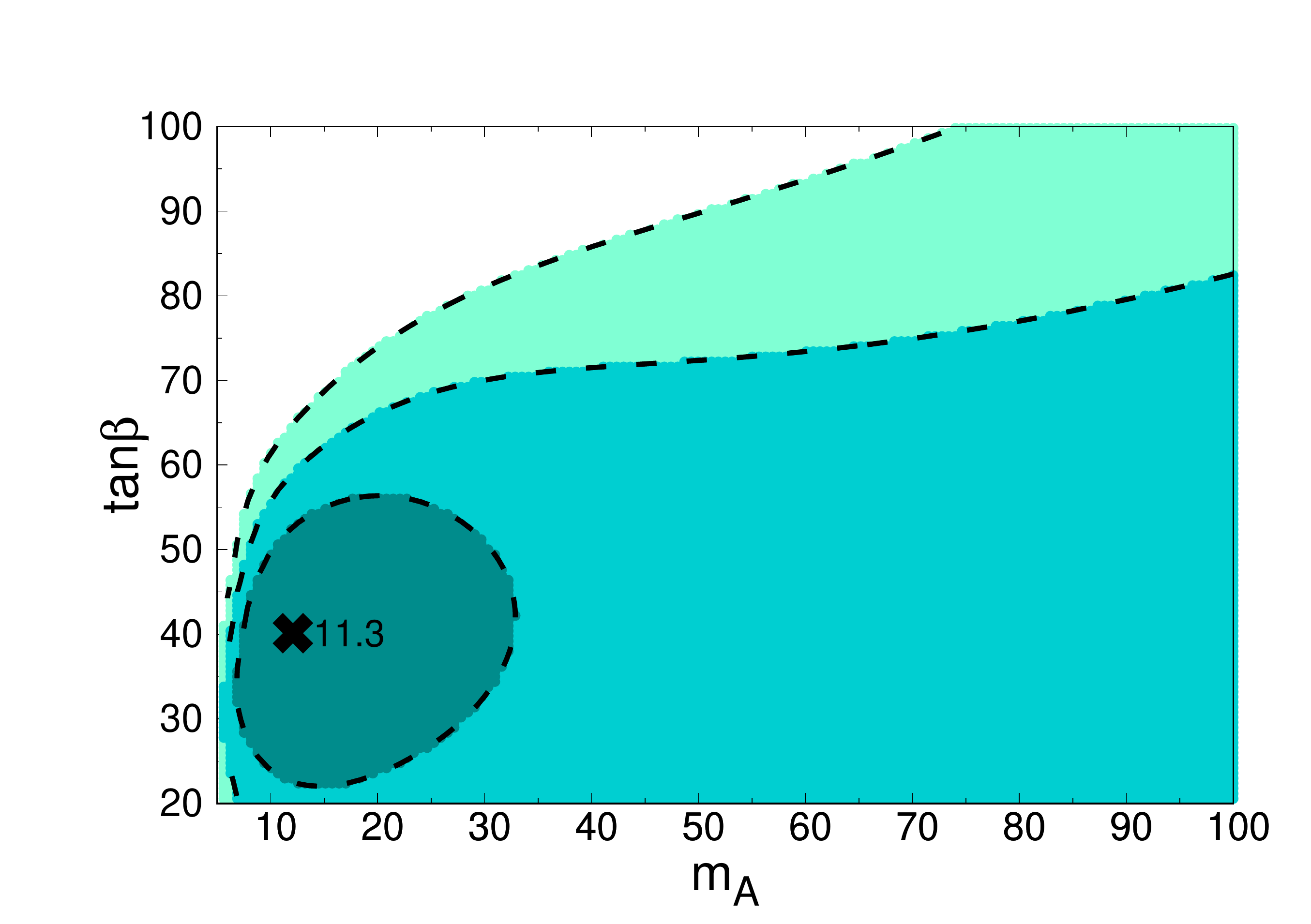}
\vspace{-0.5cm}
\caption{\small \label{fig:chi2all} 
Contours of the total $\chi^2$ fit to the muon $g-2$, $\tau$ decays, $Z$ decays, and $B_s\to \mu^+\mu^-$ for $\rho_u=0$ (left) and $\rho_u=\pi/100$ (right). The cross and the associated number 
show the best-fit point and the corresponding $\chi^2$ minimum.
As a comparison, $\chi^2_{\rm SM}=14.8$.}
\end{figure}

Fig.~\ref{fig:chi2Bsmumu} shows the $\chi^2$-contours in the $m_A$-$\tan\beta$ plane for $\overline{\rm BR}(B_s\to \mu^+\mu^-)$ with $\rho_u = 0$ (left plot) and $\rho_u = \pi/100$ (right plot).  Shapes of these contours depend sensitively on the value of 
$\rho_u$. Fig.~\ref{fig:chi2all} shows the $\chi^2$-contours for all the above-mentioned 
observables taken into account, for $\rho_u=0$ (left plot) and $\rho_u=\pi/100$ (right plot). The best fit point in $(m_A, \tan\beta)$ plane is located around  $(15~\gev, 40)$ for the whole relevant range of $\rho_u \lesssim 0.06$, which is the consistent range to the top total width measurement as we will see in the next section.
Such a set of parameters is favored in comparison with the SM ($\chi^2_{\rm SM}=14.8$) at about 1-2 $\sigma$ level. 
The $\rho_u \simeq \pi/100$ case would provide a best fit ($\chi^2 = 11.3$) mainly due to the observed $B_s \to \mu^+\mu^-$ decay branching fraction.
In summary so far, the scenario of $m_A\simeq 15$~GeV and $\tan \beta \simeq 40$ with a small mixing angle $\rho_u$ 
is preferred by muon $g-2$ without conflicts with the other observables in the up-specific VAM.

\subsection*{$D-\bar{D}$ oscillation}

In the general case when the non-zero flavor violating parameters $\psi_u$ and $\rho_u$ are allowed,
our model is constrained by more flavor observables. 
In the following, we examine the flavor violating effects on the $D$ meson mixing, 
the rare top decays, and the total top quark width.

\begin{figure}[h!]
\centering
\includegraphics[height=2.25in]{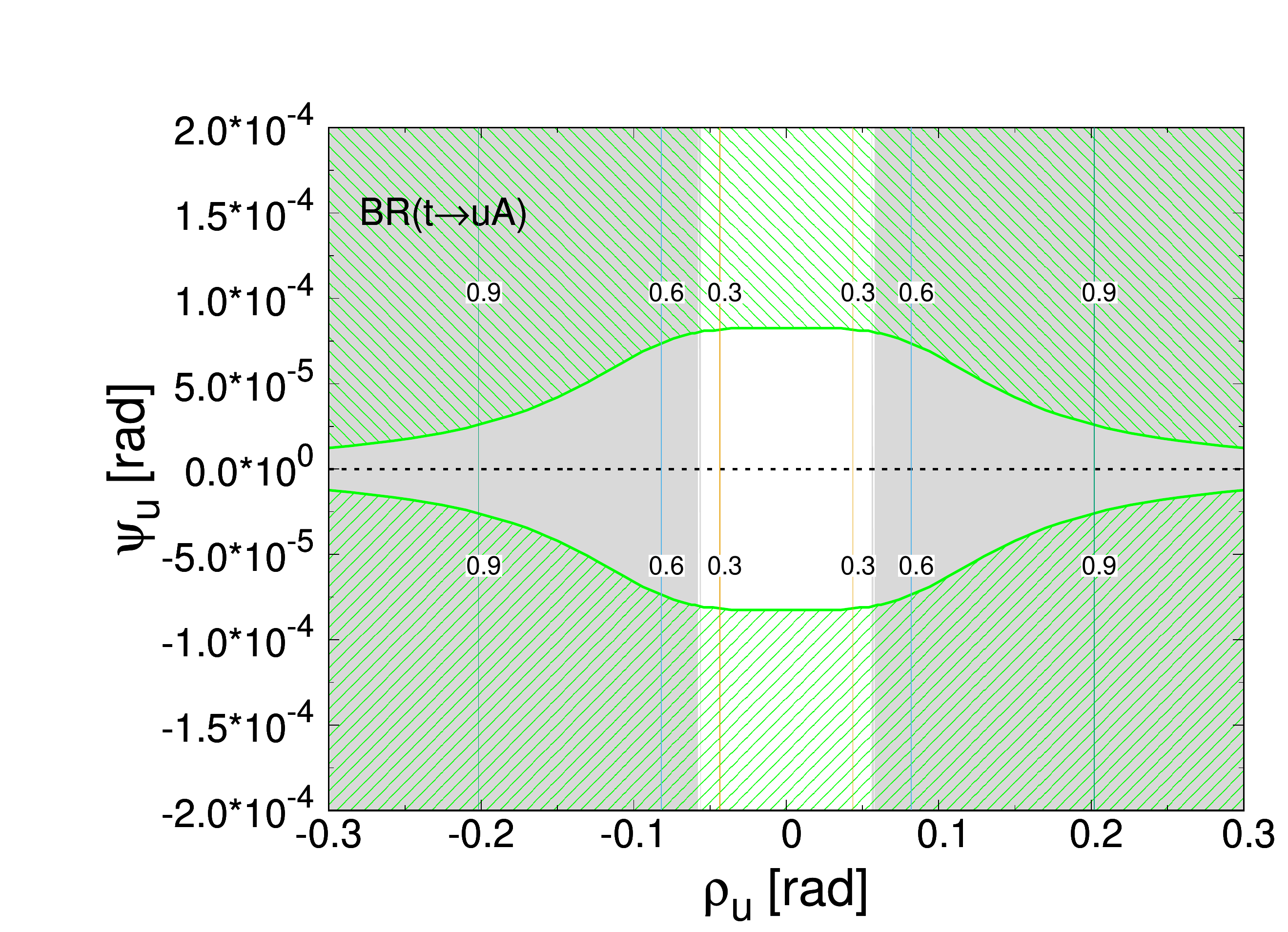}
\includegraphics[height=2.25in]{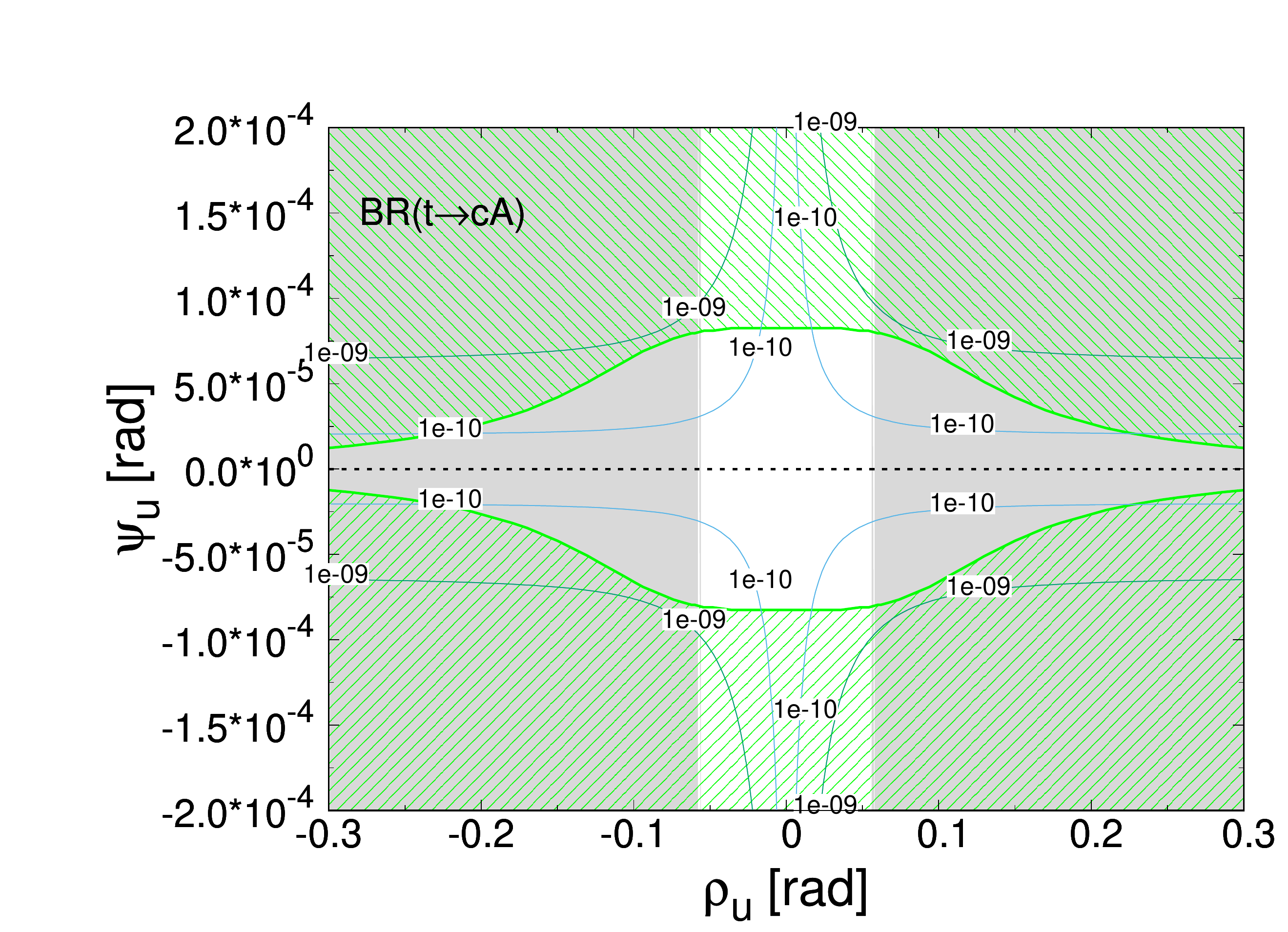}
\includegraphics[height=2.25in]{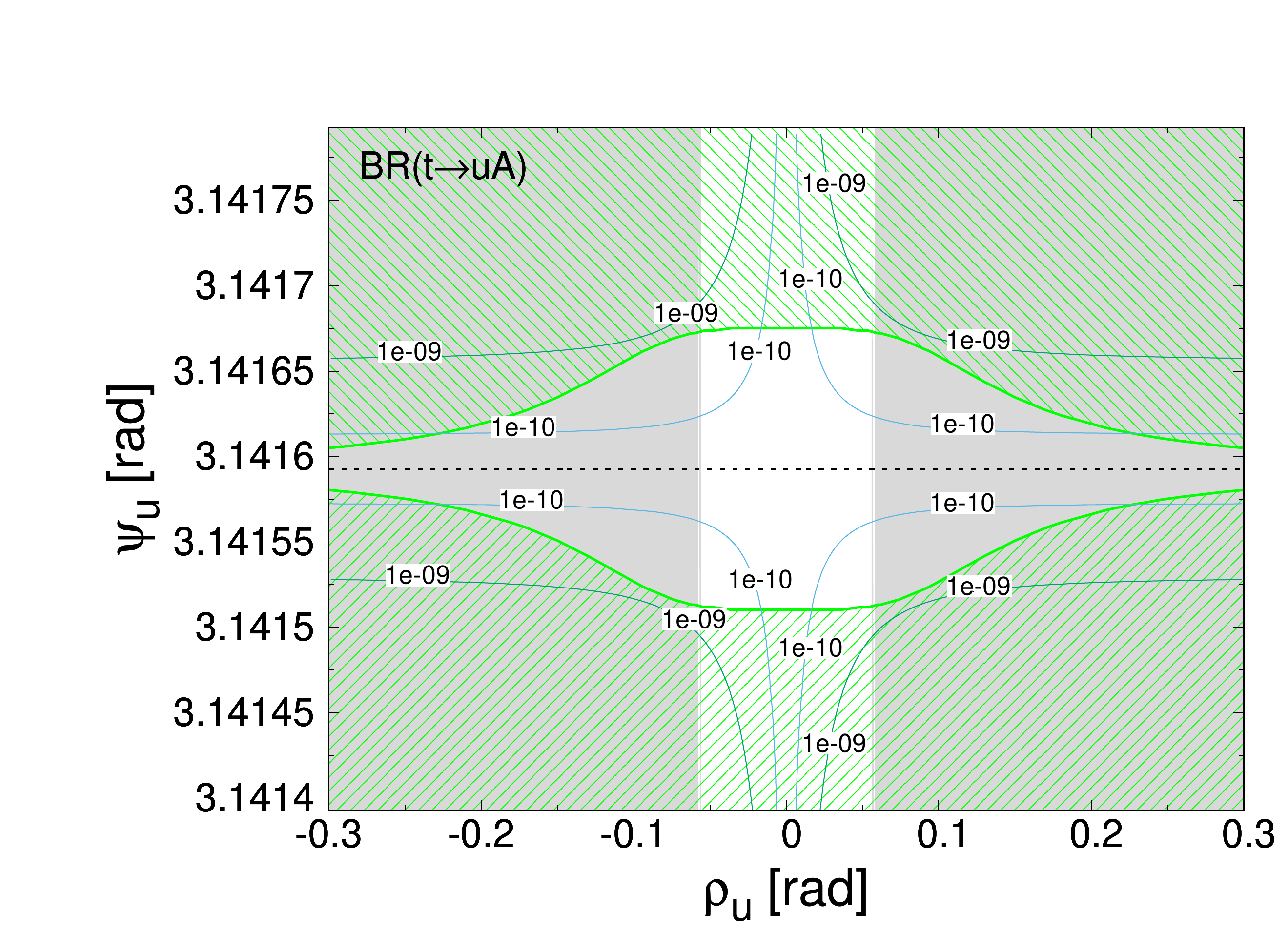}
\includegraphics[height=2.25in]{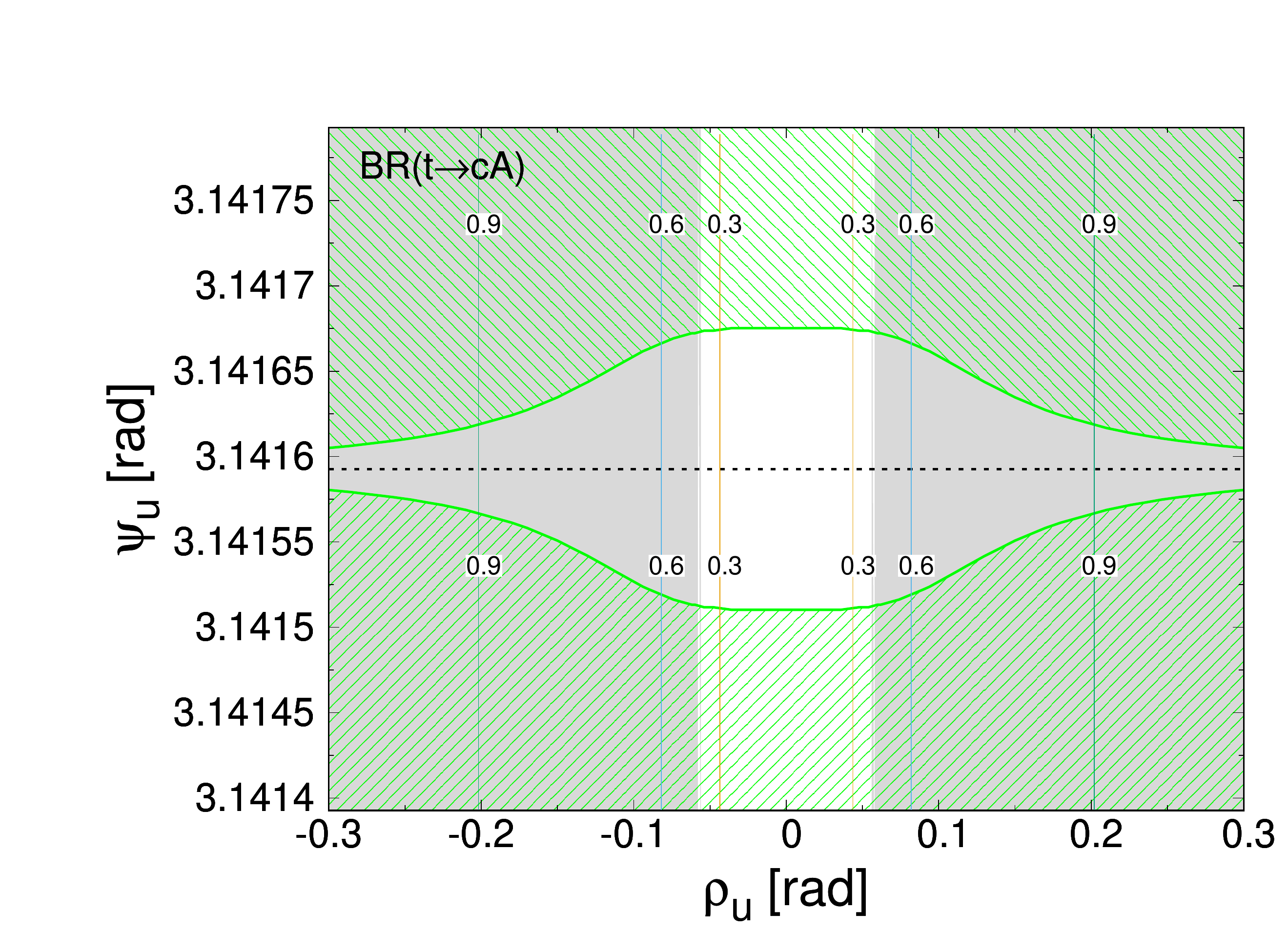}
\includegraphics[height=2.25in]{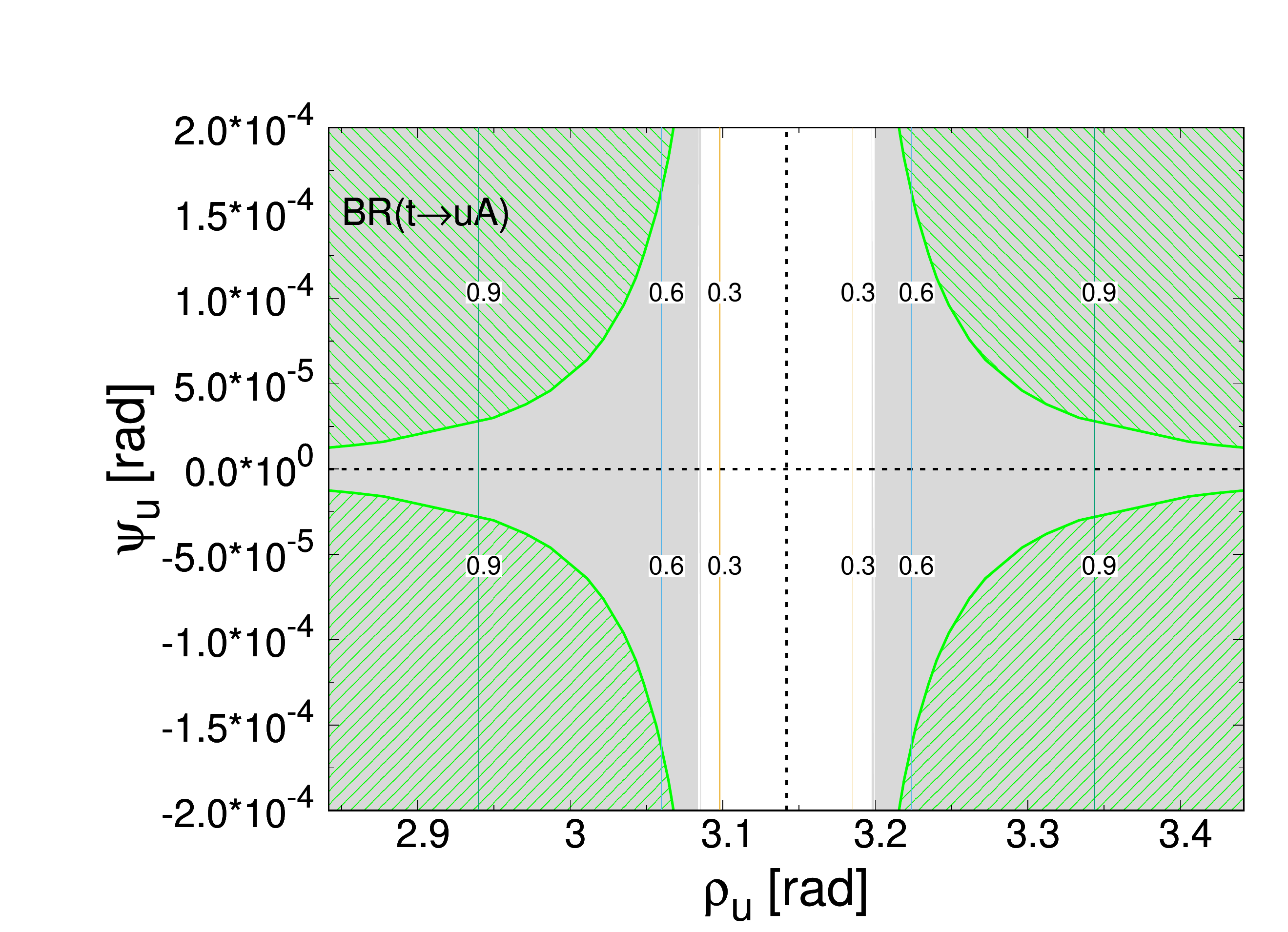}
\includegraphics[height=2.25in]{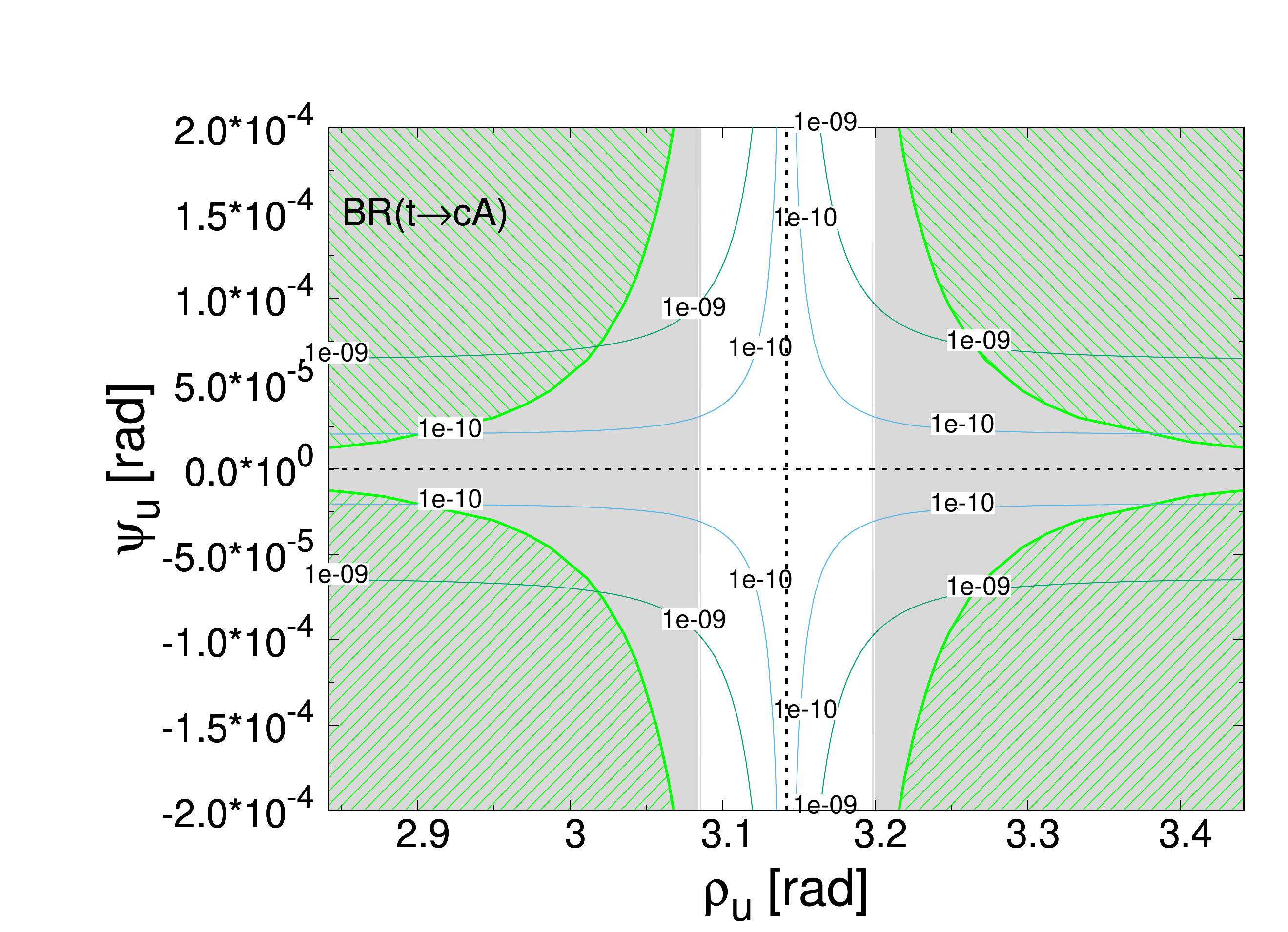}
\caption{\small \label{fig:DD} 
Allowed parameter space in the $\rho_u$-$\psi_u$ plane.  The hatched green region is excluded by the $D^0$ oscillation data at 95\% CL.
Also shown are the contours of ${\rm BR}(t \to uA)$ (left) and ${\rm BR}(t \to cA)$ (right) labeled by the values.  
The gray region is excluded by the total top-quark width measurement~\cite{CMS:2016hdd} at 95\% CL for $\tan\beta=40$. %
}
\end{figure}

The flavor changing couplings of $A$ can contribute to the $D-\bar{D}$ oscillation at both tree and loop levels.  
The tree contributions are proportional to $\xi^A_{uc}$ and described by the diagrams 
with the $s$- and $t$-channel exchanges of $A$.  The dominant loop-level contributions 
are proportional to $\xi^A_{ut} \xi^A_{ct}$ due to $A$ and $t$ running in the loop.
In the VAM, the $D^0$ oscillation imposes the following constraints~\cite{Harnik:2012pb}
\begin{eqnarray}
\label{eq25}
\mbox{tree}: &&\frac{|\xi^A_{uc}|^2}{2m^2_A}\frac{m^2_c}{v^2} \lesssim 1.6\times 10^{-13}~\gev^{-2}
\Rightarrow
\left|\frac{\tan\beta H^{uc}_u}{m_A}\right| 
\lesssim  1.1\times 10^{-4}~\gev,  \\
\label{eq26}
\mbox{loop}: && \frac{|\xi^A_{ut}\xi^A_{ct}|}{2m^2_A}\frac{m^2_t}{v^2} \lesssim 2.4\times 10^{-7}~\gev^{-2}
\Rightarrow
\left|\frac{\tan^2\beta H^{ut}_u H^{ct}_u}{m^2_A}\right| 
\lesssim  9.7\times 10^{-7}~\gev^{-2}.
\end{eqnarray}
The constraints in the $\rho_u$-$\psi_u$ plane for $\tan\beta=40$ and $m_A=15~\gev$ 
are shown in Fig.~\ref{fig:DD}, where the hatched green region is excluded at 95\% CL.
We neglect the interference effects.
It is clear that $\psi_u$ is more constrained than $\rho_u$, especially in the up-specific $(\rho_u,\psi_u)\approx (0,0)$ and charm-specific $(\rho_u,\psi_u)\approx (0,\pi)$ VAM's.  The constraints are weaker in the top-specific $(\rho_u,\psi_u)\approx (\pi, \mbox{any value})$ case. 
Even though these plots are drawn by neglecting the flavor-changing contributions from the $h,H$ bosons, 
there is virtually no change to them even when these diagrams are also included.  This is because the mass of $h,H$ are 
much heavier than $A$ in our setup and there is an additional suppression factor of 
$c_{\beta-\alpha}$ or $s_{\beta-\alpha}$ even though the $\xi^h_{ff'},\xi^H_{ff'}$ 
couplings are proportional to the same factor $\zeta_{ff'}$.

\subsection*{The $t \to hj$ constraints}

In Fig.~\ref{fig:DD}, we see that $\rho_u$ is not strongly constrained for $\psi_u \approx 0$ by the $D^0$ oscillation data.  Currently, one of the most stringent constraints comes from the top FCNS decays of $t \to h j$, where $j=u,c$ and is given by~\cite{Chiang:2017fjr} 
\begin{eqnarray}
\label{eq27}
| c_{\beta-\alpha} (\tan\beta + \cot\beta) \sin\rho_u|
\lesssim 0.26\,.
\end{eqnarray}

The exclusion region generally depends on the value of $\tan\beta$ and $c_{\beta-\alpha}$,
which is empirically found to be close to zero from 125-GeV Higgs coupling measurements.  
For example, with $c_{\beta-\alpha}=0.3$ and $\tan\beta=40$, the constraint reads $\sin\rho_u \lesssim 0.02$.
In the alignment limit $c_{\beta-\alpha}=0$, however, these FCNS top decays vanish identically in the up-type VAM.

\subsection*{The $t \to Aj$ constraints from top quark total width}

A robust constraint on the FCNS couplings independent of $c_{\beta-\alpha}$ can be obtained 
from the model-independent top width measurements~\cite{CMS:2016hdd}, 
where the total top width is reported not to exceed $2.5$~GeV at 95\% CL.  
On the other hand, the VAM's predict the rare top decays of $t\to uA$ and $cA$.
We demand that the total 
top decay width, including the standard partial width 
$\Gamma_{t\to bW} = 1.41$~\gev, be below the the upper bound; that is, 
$\Gamma_{t,{\rm tot}} \equiv \Gamma_{t\to bW} + \Gamma_{t\to uA/cA} \leq 2.5~\gev$ with
\begin{eqnarray}
\Gamma_{t \to uA/cA }=\frac{G_F m^3_t}{64\pi \sqrt{2}}\sin^2\rho_u (\tan\beta+\cot\beta)^2 
\left(1-\frac{m^2_A}{m^2_t} \right)^2 .
\end{eqnarray}
The condition $\Gamma_{t \to uA/cA } \leq 1.1~\gev$, corresponding to $BR(t \to uA/cA) \lesssim 40\%$, 
is translated to $|\rho_u| \lesssim 0.06$ for $\tan\beta=40$ and $m_A=15~\gev$.
Note that it currently provides the most stringent constraint on $\rho_u$.
The exclusion region is shown by the gray region in Fig.~\ref{fig:DD}.

\subsection*{The $h\to AA$ decay}

When the $A$ boson is lighter than $m_h/2$, the decay of $h \to AA \to 4\tau$ is kinematically allowed.  
Although there is a CMS study constraining $h \to AA \to 2\tau 2b$ $\lesssim 2-10$\%~\cite{Sirunyan:2018pzn}, we cannot find the constraint on $h \to AA \to 4\tau$. 
Thus, we consider the branching ratio of exotic Higgs decays, which is currently bounded to ${\rm BR}(h \to AA)\lesssim 20\%$~\cite{Curtin:2013fra}.  
Using the partial width formula
\begin{align}
\Gamma(h \to AA)=\frac{1}{32 \pi}\frac{\lambda^2_{hAA}}{m_h}\sqrt{1-\frac{4m^2_A}{m^2_h}}
~,
\end{align}
we find that the trilinear scalar coupling should satisfy
$
|\lambda_{hAA}|\lesssim 3.7~{\rm GeV}
$.
In the lepton-specific 2HDM with large $\tan\beta$, $\lambda_{hAA}$ can be expressed 
in terms of neutral Higgs boson masses and lepton Yukawa couplings as~\cite{Chun:2015hsa}
\begin{align}
\lambda_{hAA}\simeq 
\frac{2m^2_A+\xi^h_{\tau\tau}s_{\beta-\alpha}m^2_h
   -(s^2_{\beta-\alpha}+\xi^h_{\tau\tau}s_{\beta-\alpha})m^2_H}{v}\,.
\end{align}
From the first two terms in the numerator, the natural size of $\lambda_{hAA}$ is about 60~GeV. 
Therefore, ${\cal O}(10\%)$ fine tuning is required for a cancellation between the first two terms 
and the third term in the numerator, in order to satisfy the experimental constraint on the $h\to AA$ decay.

Under the constraints of Higgs data: $s_{\beta-\alpha}\simeq 1$ and $|\xi^h_{\tau\tau}|\simeq 1$, we can consider two possible cases.  
The first case corresponds to the right-sign limit ($\xi^h_{\tau\tau}\to +1$). 
However, in the large $\tan\beta$ limit, the bound of $m_H<250$~GeV is required 
by the perturbativity condition of the $\lambda_{hAA}$ coupling.  The charged Higgs 
mass is then forced to be light due to the condition $m_{H^\pm}\simeq m_H$ without conflicting with the electroweak precision measurements. 
Consequently, the region associated with the right-sign limit is ruled out due to the direct search of the charged Higgs boson at the LHC.
The second case corresponds to the wrong-sign limit ($\xi^h_{\tau\tau}\to -1$), 
where $m_H$ can be arbitrarily large and, therefore, $\lambda_{hAA}$ can be fine-tuned to vanish.
The solution for an exact cancellation in $\lambda_{hAA}$ in the wrong-sign limit is
\begin{align}
c_{\beta-\alpha}\simeq \frac{1}{\tan\beta}\left(2+\frac{m^2_h-2m^2_A}{m^2_H}\right)\,,
\end{align}
for $m_H \gtrsim 300$~GeV.  This is almost independent of $m_A$ provided $m_H \gg m_A$, as is the case considered here.

\subsection*{Single $A$ production at the LHC}

In the up-specific VAM, the $A$ production at the LHC through the process $p p \to u \bar{u} \to A$ 
is enhanced by $\tan\beta$.
For example, for $m_A=20$~GeV, the production cross section at the 13-TeV LHC reaches
\begin{equation}
\sigma(pp \to u \bar{u} \to A)\simeq 0.8 \left(\frac{\tan\beta}{40}\right)^2\,\, {\rm pb}. 
\end{equation}
It is, however, too small to detect through the di-muon resonance signal with the current integrated luminosity~\cite{dimuon}.  
\bigskip

In this section, we have examined various current constraints on the up-type VAM.
As a short summary, we have found that the up-specific VAM with $m_A \sim 15~\gev$ and $\tan\beta \sim 40$ 
can provide a good explanation for the deviation found in $(g-2)_\mu$, 
while the charm-specific VAM is not favored in this regard.
There are no conflict with any flavor observables and experimental measurements considered in this section, as long as the mixing angles $\rho_u$ and $\psi_u$ are small. 
Moreover, taking non-zero $\rho_u$ at the order of $\pi/100$ offers a better fit to the $B_s \to \mu\mu$ branching fraction.  This could lead to interesting phenomenology.

\section{Smoking gun signatures at LHC}
\label{sec:quark_FCNC}

\subsection{Rare top decay $t \to u + A$}

In this section we focus on the up-specific VAM and first consider the most promising 
signature $t \to u + A$ at the LHC.  The analysis shown in this section 
can be readily applied to the charm-specific case with $t \to c + A$, 
although as seen in the previous section this case is slightly disfavored by the $(g-2)_\mu$ observation.
\medskip 

In the VAM with $\psi_u = 0$, the $t \to u+A$ decay helicity amplitudes ${\cal M}_{h_t,h_u}$ in the limit of $m_u= 0$ are given by 
\begin{align}
\begin{split}
&
{\cal M}_{++} = {\cal M} \cos\frac{\theta}{2} ~,~~~ 
{\cal M}_{-+} =  -{\cal M} \sin\frac{\theta}{2} ~,~~~
{\cal M}_{+-} =  {\cal M}_{--} = 0 ~,
\\
&
\mbox{where }~   {\cal M} \equiv \frac{\sqrt{2m_t E_u}\, m_t \sin\rho_u}{2v}(\tan\beta+\cot\beta).
\end{split}
\end{align}
The angle $\theta$ is between the up-quark 3-momentum and the top spin and the up-quark energy $E_u=\frac{m_t}{2}(1-\frac{m^2_A}{m^2_t})$ in the top rest frame.  The direction of up-quark emitted from the top quark is preferentially aligned with the polarization of the top quark.  The partial decay width of $t \to uA$ is easily computed as 
\begin{equation}
\Gamma_{t \to uA}=\frac{G_F m^3_t \sin^2\rho_u}{64\pi\sqrt{2}}
(\tan\beta+\cot\beta)^2
\left(1-\frac{m^2_A}{m^2_t} \right)^2\,,
\end{equation}
which is not suppressed even in the alignment limit, in contrast to the $t \to uh$ partial decay width that is suppressed by $c_{\beta-\alpha}$.  The branching ratio assuming $\Gamma_{t\to uA} \ll \Gamma_{t\to bW} $ is, with $r_{a/b}=m_a^2/m_b^2$:
\begin{eqnarray}
{\rm BR}(t \to uA) =
\frac{(1-r_{A/t})^2}{8(1-r_{W/t})^2(1+2r_{W/t}) |V_{tb}|^2} \sin^2\rho_u (\tan\beta+\cot\beta)^2 
\simeq 0.14 \rho_u^2 \tan^2\beta.
\end{eqnarray}
Therefore, for $\tan\beta=40$, a ${\cal O}(0.01)$ of $\rho_u$ would provide the ${\cal O(\%)}$ branching ratio:
$\left(\rho_u/0.01\right)^2 \times 2.24\%$.

\subsection{Current constraints from existing searches}

As top quarks are copiously produced in pairs at the LHC, one should be able to constrain or search for this rare FCNS $t\to uA$ decay.  
Nevertheless, we cannot find a dedicated experimental study searching for this specific rare decay in the literature other than several theoretical studies~\cite{Kao:2011aa,Chen:2013qta,Altunkaynak:2015twa}.
We consider the main production mechanism of the pseudoscalar $A$ 
as $pp \to t\bar{t} \to (b\ell\nu)(uA) \to (b\ell\nu)(u\tau^+\tau^-)$, which involves one $b$-jet from the standard top decay.
The relevant analyses indirectly constraining this process and ${\rm BR}(t \to u A)$ at the LHC are the light pseudoscalar Higgs boson searches in association with a $b\bar{b}$ pair~\cite{CMS:2015mca, Khachatryan:2015baw,Sirunyan:2017uvf}, 
with several theoretical efforts being made to improve the sensitivity~\cite{Goncalves:2016qhh,Banerjee:2017wxi,Bernon:2014nxa}.
Among them we find that the CMS analysis at 8~TeV~\cite{Khachatryan:2015baw} currently 
provides the most stringent constraint. The CMS has shown that the sensitivity using $A \to \mu\mu$ 
is comparable but  weaker than that using $A \to \tau\tau$, 
assuming ${\rm BR}(A\to\tau\tau)/{\rm BR}(A\to \mu\mu) =(m_\tau/m_\mu)^2$~\cite{Sirunyan:2017uvf}.
There are also searches using the di-tau channel on the 13-TeV data~\cite{Aaboud:2017sjh}, 
although they focus on the case where the new bosons are heavy and only show a limit of $m_A = 300$~GeV at lightest.

We follow the most stringent 8-TeV CMS analysis~\cite{Khachatryan:2015baw} and re-interpret it to constrain our model.  For the signal analysis, we use {\tt MadGraph5+Pythia8}~\cite{Alwall:2014hca,Sjostrand:2007gs} for event generation and {\tt Delphes3}~\cite{delphes3} for detector simulation. 
For jet reconstruction, we rely on the {\tt FastJet} package~\cite{Cacciari:2011ma} and use the anti-kT algorithm with the standard jet size $R=0.5$.
The data include three channels: $e\mu, e\tau_h$, and $\mu \tau_h$, where $\tau_h$ denotes a tau lepton decaying hadronically, and their respective selection cuts are summarized as follows: 
\begin{itemize}
\item $\mu\tau_h$ channel: 
exactly one $\mu$ and one $\tau_h$ with opposite charges: \\
$p_{T,\mu}>18$~GeV, $|\eta_\mu| < 2.1$ and $p_{T,\tau_h}>22$~GeV, $|\eta_{\tau_h}| < 2.3$,\\
$\Delta R(\mu, \tau_h) >0.5$, $M_T(p_{T,\mu}, \vec{p}\!\!\!/_T) < 30~\gev$,

\item$e\tau_h$ channel:  
exactly one $e$ and one $\tau_h$ with opposite charges: \\
$p_{T,e}>24$~GeV, $|\eta_\mu| < 2.1$ and $p_{T,\tau_h}>22$~GeV, $|\eta_{\tau_h}| < 2.3$,\\
$\Delta R(e, \tau_h) >0.5$, $M_T(p_{T,e}, \vec{p}\!\!\!/_T) < 30~\gev$,

\item $e\mu$ channel: 
exactly one $\mu$ and one $e$ with the opposite charge: \\
\ \ \ \  \ \ \ \ \ ($p_{T,\mu}>18$~GeV, $p_{T,e}>10$~GeV) or
($p_{T,\mu}>10$~GeV, $p_{T,e}>20$~GeV), \\
$|\eta_\mu| < 2.1$ and $|\eta_{e}| < 2.3$\\
$\Delta R(e, \mu) >0.5$, $M_T(p_{T,e}+p_{T,\mu}, \vec{p}\!\!\!/_T) < 25~\gev$,
$P_\zeta - 1.85 P_\zeta^{\rm vis} > -40~\gev$,
\end{itemize}
where $M_T(p_T,  \vec{p}\!\!\!/_T))= \sqrt{p_T p\!\!\!/_T(1 - \cos\phi)}$ and $\phi$ is the azimuthal 
angle difference between the momentum and the missing transverse momentum.
The definitions of $P_\zeta, P_\zeta^{\rm vis}$ can be found in Ref.~\cite{Khachatryan:2015baw}.
In addition to the above selection cuts, events in all the channels are required to have at least one $b$-tagged jet with $p_{T,b}>20~\gev$ and $|\eta_b|<2.4$.
For the hadronic $\tau$ tagging, we take a simpler algorithm described in Ref.~\cite{Papaefstathiou:2014oja}
for the Cambridge/Aachen (C/A) jets with $R=0.5$ and call a $\tau$ tag when the following conditions are satisfied: 
\begin{itemize}
\item define $j_{\rm core}$ by drawing a cone with a smaller radius $R=0.1$ centered at the jet, 
and require no tracks with $p_{T,\rm track}>1$~GeV to lie in the annulus between $0.1<R<0.5$;
\item the hardest track in $j_{\rm core}$ satisfies $p_T> 5$~GeV; and
\item $f_{\rm core} > 0.95$, where $f_{\rm core} \equiv \sum_{R<0.1} E_T^{\rm calo}/\sum_{R<0.2}E_T^{\rm calo}$, the fraction of jet energy deposited in the jet core.
\end{itemize}
For the selected events, we consider the possible values of $m_{\tau\tau}$ consistent with the kinematics 
of the two visible tau decay products, and take the minimum value as $m_{\tau\tau}$.
We can then set an upper limit on the cross section at 95~\% CL. from the combined $m_{\tau\tau}$ distribution of all three channels.
\medskip 

\begin{figure}[h!]
\centering
\includegraphics[height=2.7in]{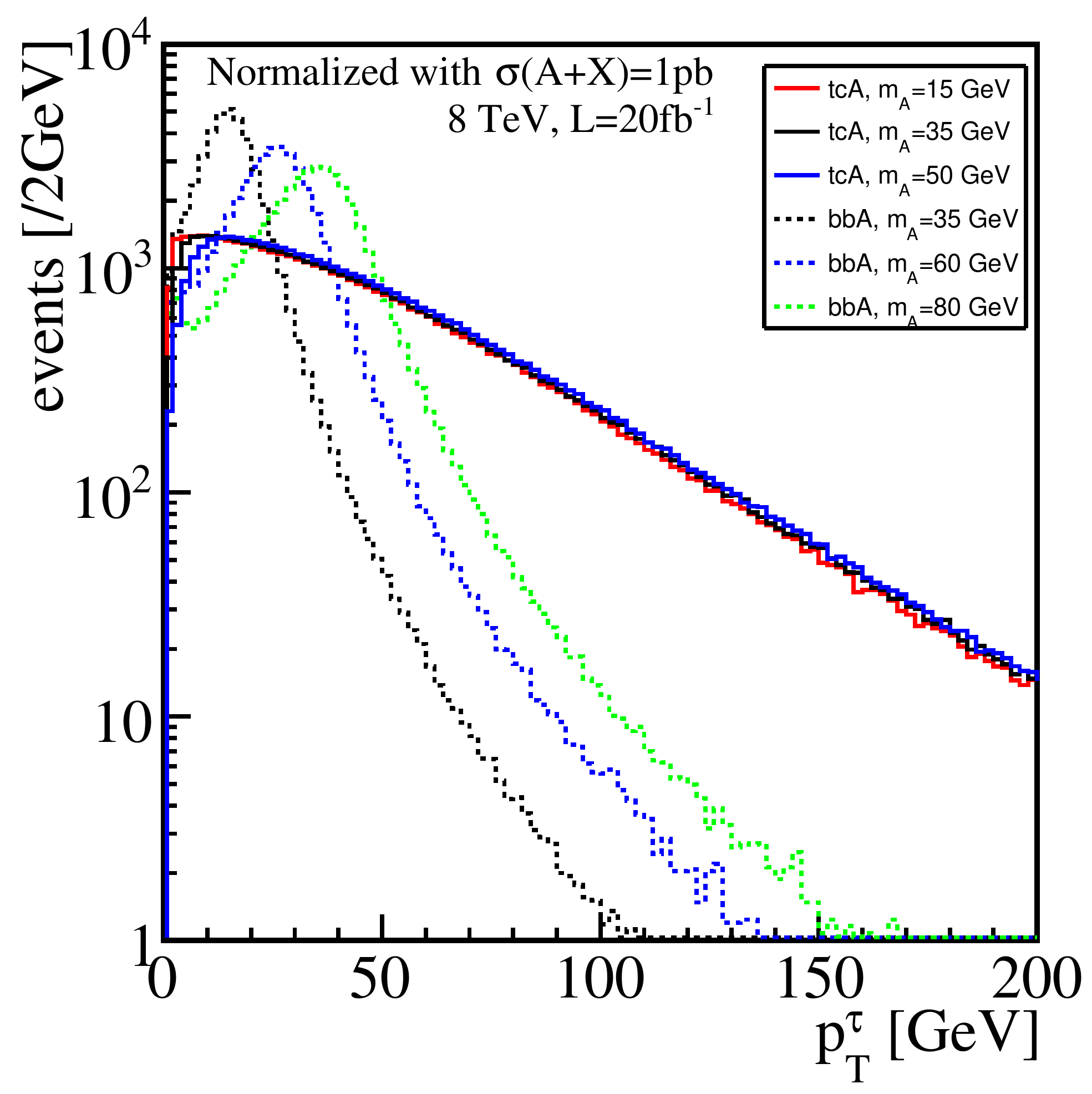}
~~
\includegraphics[height=2.7in]{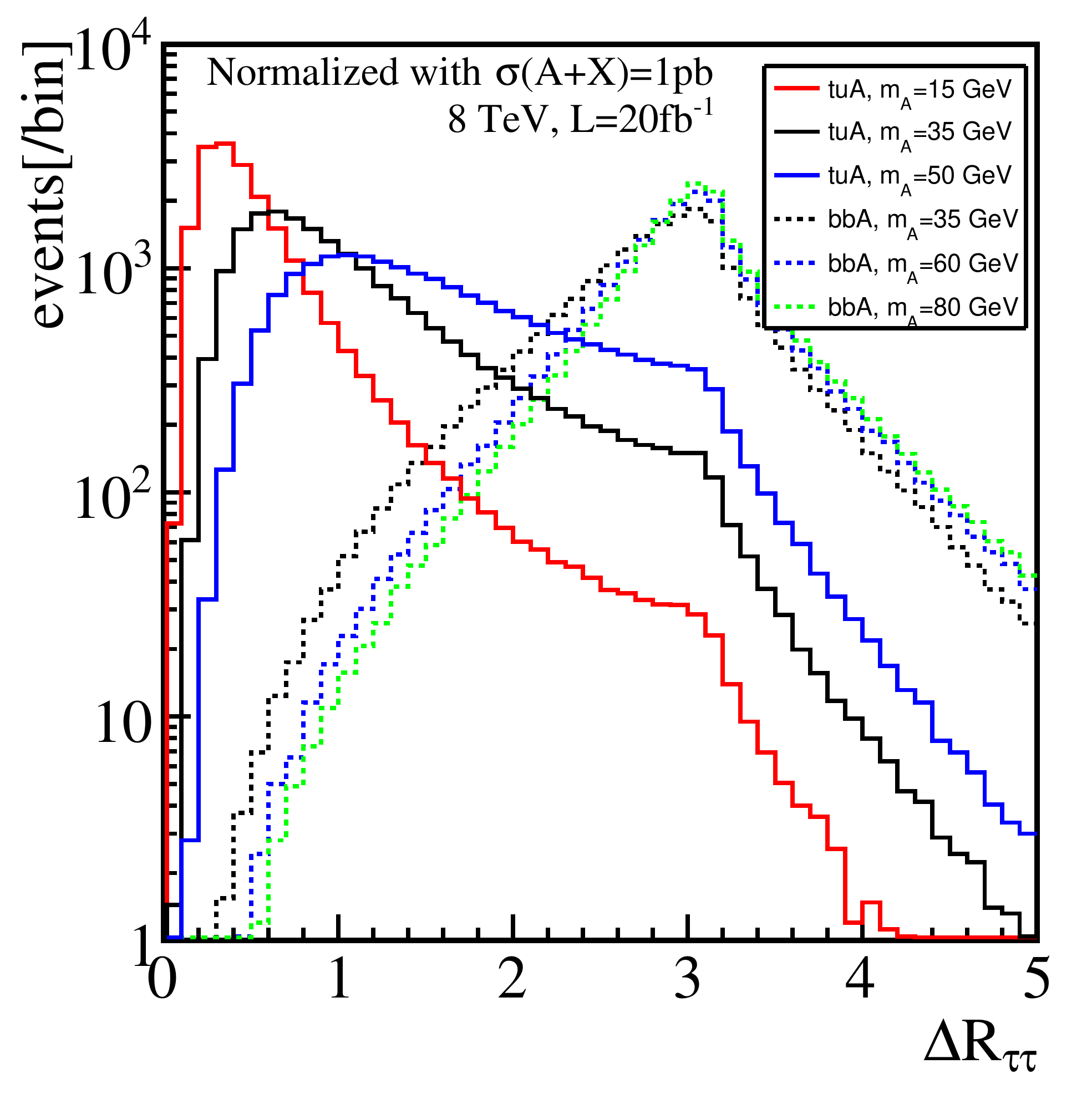}

\includegraphics[height=2.7in]{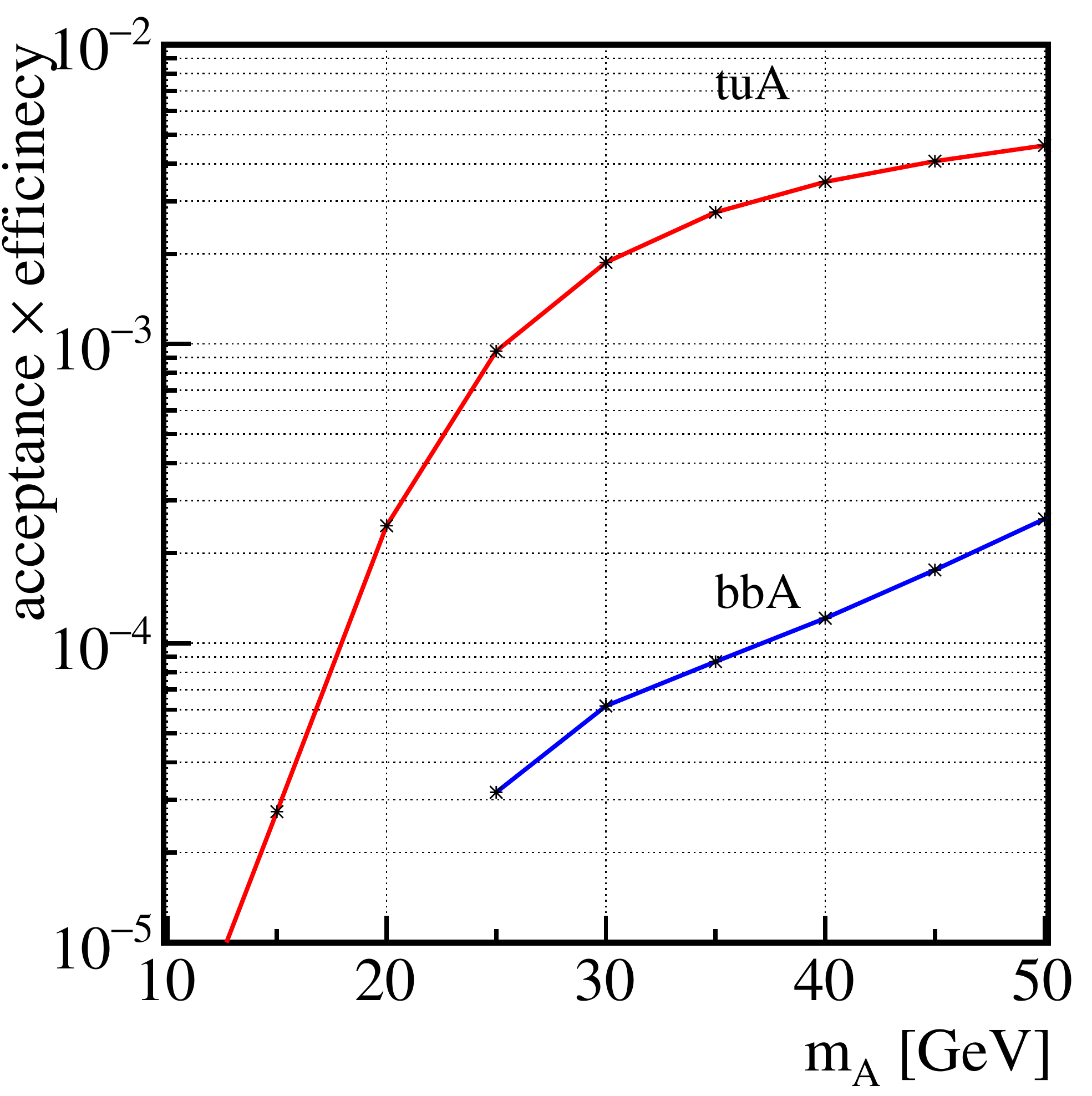}
~~
\includegraphics[height=2.7in]{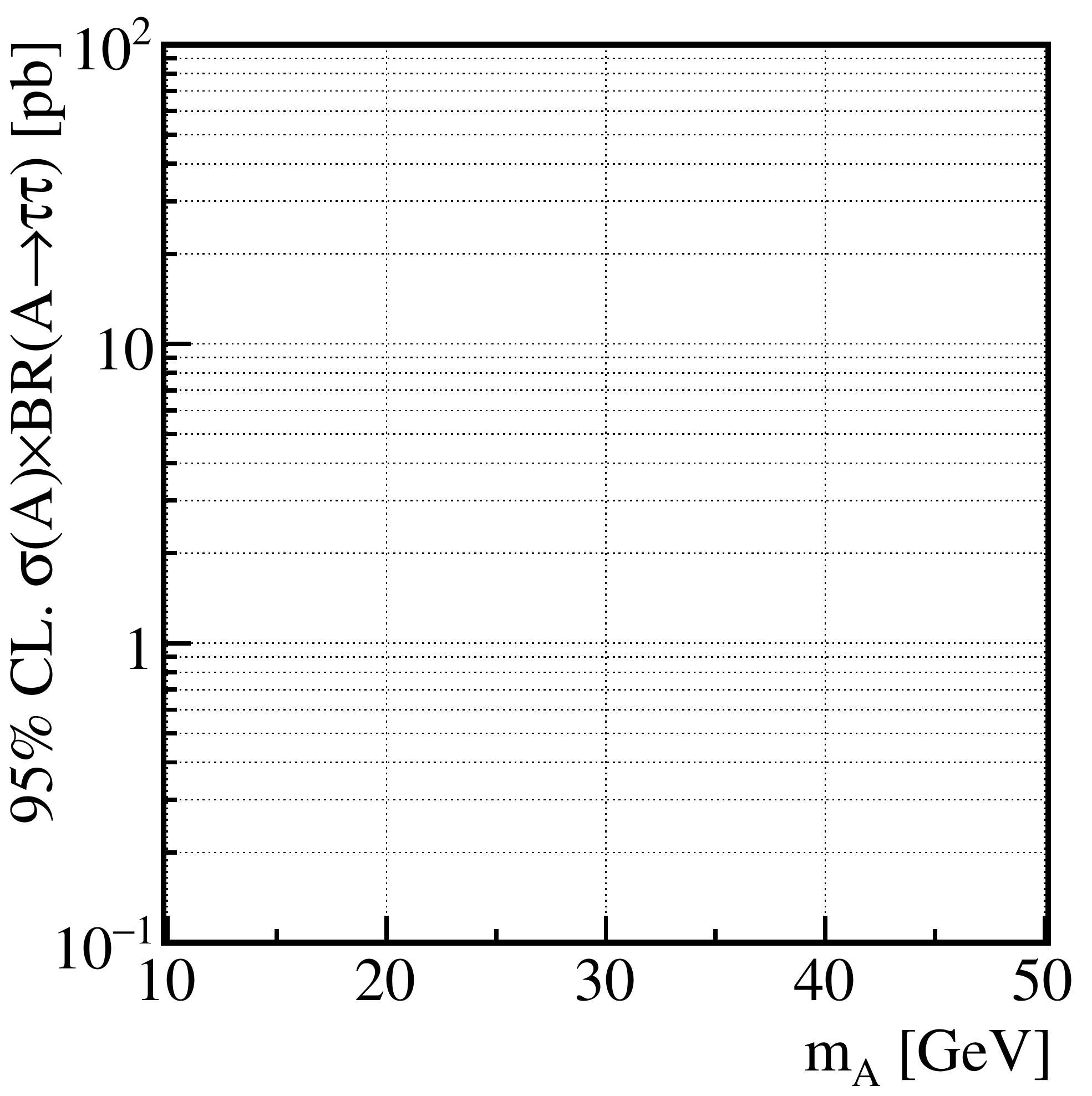}
\caption{The upper left and upper right plots show respectively the $p_T$ and $\Delta R_{\tau\tau}$ distributions of $\tau$ leptons for several $m_A$ values in the $bbA$ associated production and $tuA$ (from $t\bar{t}$) production.  The lower left and lower right plots show respectively the acceptance times efficiency and the cross section upper bounds on the $bbA$ (blue curve) and $tuA$ (red) channels as a function of $m_A$.
} 
\label{fig:taupt}
\end{figure}

There are two important differences between  the $A\to \tau\tau$ decay from $b\bar{b}A$ production 
and that from the $t\to uA$ decay in $t\bar{t}$ production (in this section, we shall refer to them as $bbA$ and $tuA$, respectively).
One is that the $p_{T,\tau}$ distribution is softer in the $bbA$ sample when $m_A$ is small, whereas $p_{T,\tau}$ in $tuA$ 
sample is harder and practically independent of $m_A$, especially in the tail.
This is seen in the upper-left plot of Fig~\ref{fig:taupt}, where the normalized $p_{T,\tau}$ distributions 
in the selected $bbA$ and $tuA$ samples are shown.
The other is that the $\Delta R(\tau_1, \tau_2)$ distribution is peaked at $\sim \pi$ in $bbA$ samples whereas 
it is peaked at $\sim 0$ for $tuA$, and more prominent when $m_A$ becomes smaller.  
The normalized $\Delta R_{\tau\tau}$ distributions are shown in the upper-right plot of Fig.~\ref{fig:taupt}.
As a result, the efficiencies of the selection cuts are rapidly falling as $m_A$ gets smaller in both cases of $bbA$ and $tuA$ but for different reasons.  The resulting acceptance and efficiency as a function of $m_A$ is shown in the lower-left plot of Fig.~\ref{fig:taupt}.
As $\tau$-tagging efficiency rapidly falls as $p_{T,\tau}$ decreases, so is the lepton acceptance from the leptonic tau decay.  The final efficiency for the $bbA$ production diminishes when $m_A$ becomes small.
On the other hand, the acceptance of the $tuA$ channel is relatively high.  For example, it is about 30 times higher than the $bbA$ channel for $m_A=30$~GeV.
Taking this efficiency difference into account, we can re-interpret and apply the constraints obtained for the $bbA$ production to the $tuA$ production.
The resulting upper bound on the $tuA$ production cross section at the 95~\% CL is shown as a function of $m_A$ in the lower-right plot of Fig.~\ref{fig:taupt}.
Assuming $\sigma(t\bar{t}) = 250$~pb at LHC 8~TeV and $BR(t\to uA)$ is sufficiently small, we can translate the upper bound on the cross section to that on the branching ratio and obtain roughly $BR(t \to uA) \lesssim 0.3$~\% for $m_A\ge 25$~GeV.
For $m_A<25$~GeV, the $tuA$ acceptance is exponentially falling due to the $\Delta R_{\ell\ell'}>0.5$ cut.
Based on our estimate, $BR(t \to uA) \lesssim 10$~\% would be allowed for $m_A=15~\gev$. 
Since the CMS analysis does not show the results for $m_A<25$~GeV, we apply in this range the same upper bound on the cross section times the acceptance given at $m_A=25$~GeV.
We consider our results in that range as a conservative estimate because 
we expect smaller SM background contributions for the appropriate signal region.

\subsection{Searches using di-tau tagging}
In the previous section we have shown that the existing searches become insensitive 
for the light $A$ region, of most interest to us for explaining the muon $g-2$ in our model.  
In this section we propose an effective way to probe the region, and provide a rough estimate 
of the expected sensitivity using the current and future data at the LHC.

\begin{figure}[h!]
\centering
\includegraphics[height=2.7in]{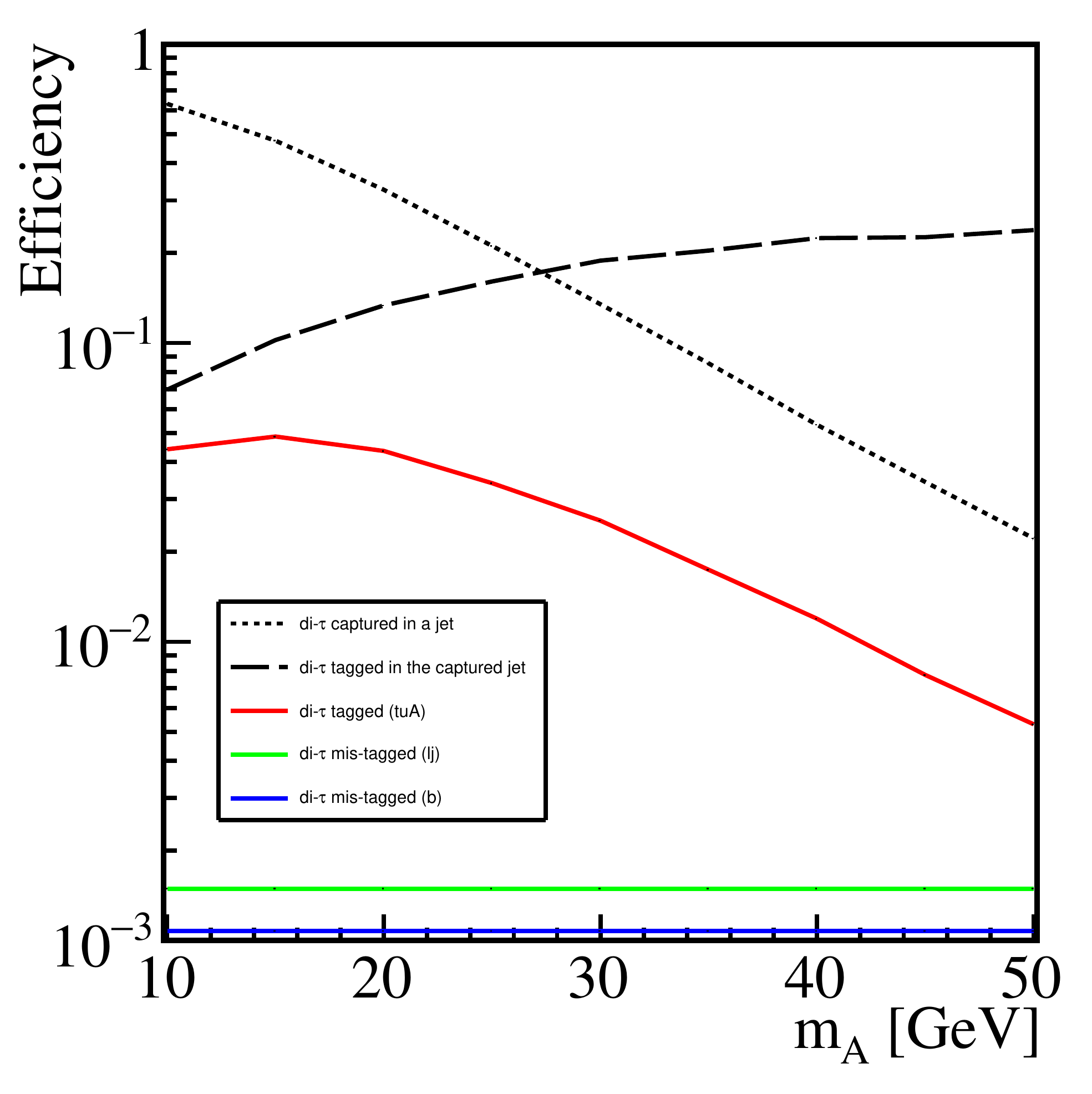}
~~
\includegraphics[height=2.7in]{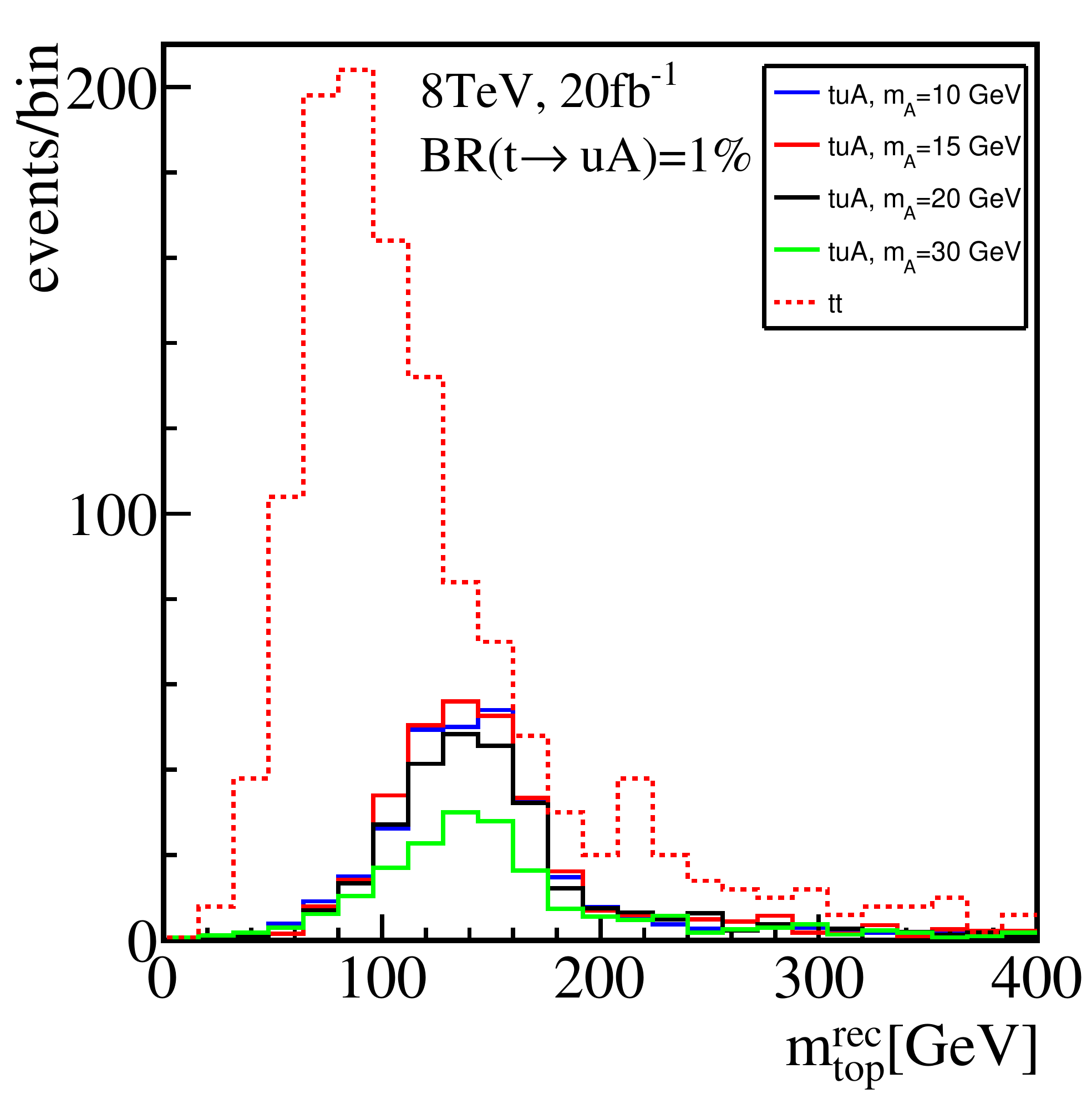}

\includegraphics[height=2.7in]{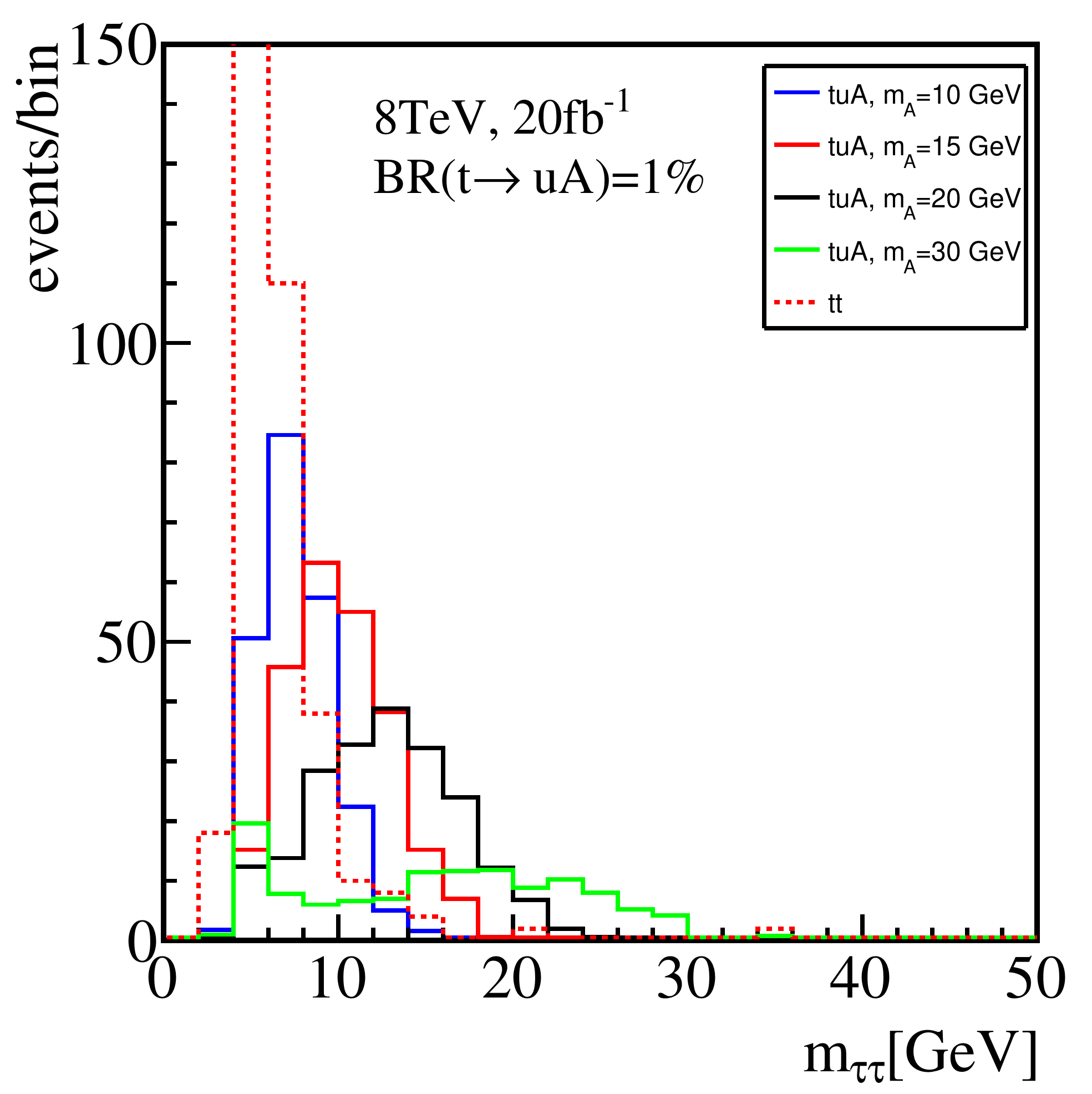}
~~
\includegraphics[height=2.7in]{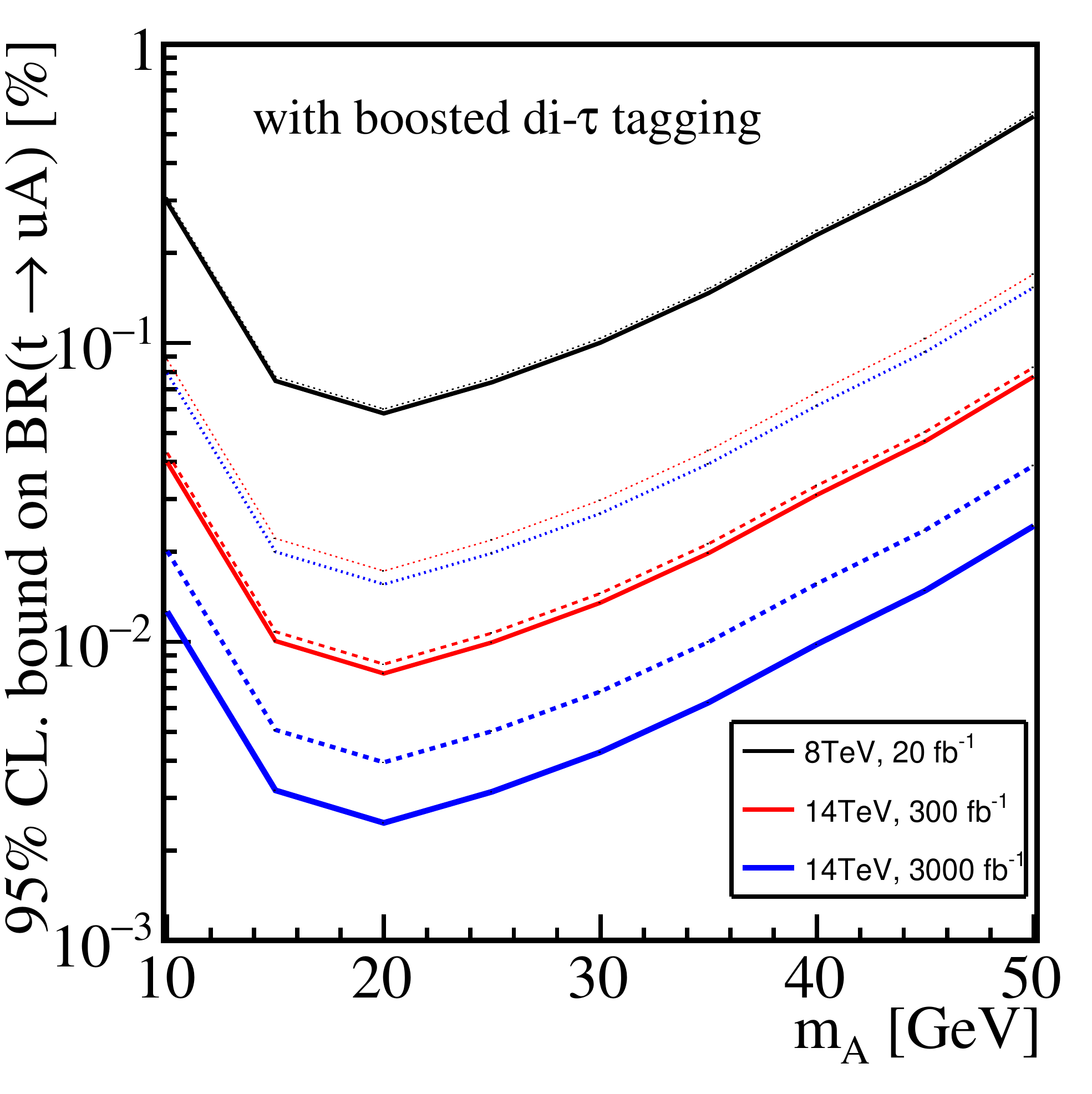}
\caption{The upper left plot shows the efficiency of the di-$\tau$ tagging algorithm described in the main text.  The upper right and lower left plots show respectively the reconstructed top mass distributions and di-$\tau$ jet mass distributions for the signal and background events.
The lower right plot gives estimated current and future sensitivities on the $BR(t \to uA \to u\tau^+\tau^-)$ with 5\%, 1\%, 0\% systematic uncertainty for the background events.
} 
\label{fig:ditau}
\end{figure}

One of the main reasons for the sensitivity to present a sharp drop in the light $m_A$ region
is due to the $\Delta R_{\ell\ell'}>0.5$ cut on the reconstructed leptons and taus.
For a light $A$, it is boosted from the top decay with the decay products (a di-tau pair from $A$) being collimated.  They are difficult to discriminate and hence naturally captured as one object.
We develop the di-tau tagging algorithm following the idea of {\it mutual isolation} proposed in the  Ref.~\cite{Katz:2010iq} as follows:

\begin{itemize}
\item find a jet using the C/A algorithm with $R=0.5$;
\item define two exclusive sub-jets $j_{1}$ and $j_{2}$ using calorimeter towers in the jet;
\item define tau-candidates $j_{1}^{\rm core}$, $j_{2}^{\rm core}$ by drawing a cone of $R_{\rm core}=0.1$ around each sub-jet;
\item for both $i=1, 2$, require that once the activities (tracks and calorimeters) in $j_{i}^{\rm core}$ are removed, the remaining activity in the original jet satisfy the tau-tagging criteria, 
with $f_{\rm core}^{\rm cut}=0.9$; and
\item the hardest tracks in $j_1^{\rm core}$ and $j_2^{\rm core}$ are oppositely charged. 
\end{itemize}

Based on the algorithm described above, the resulting tagging efficiency is shown in the upper left plot of Fig.~\ref{fig:ditau}.
Note that the tagging efficiency depends on the definition: since the efficiency definition is in general the number of jets tagged divided by the number of preselected jets, 
there are several possibilities to choose the preselected number as the denominator.
First, the efficiency of both visible decay products from the taus in an $A$ decay are captured in a jet is shown by the dotted curve as a function of $m_A$, which rapidly drops as $m_A$ gets larger from 60\% to 3\%.
The tagging efficiency of this algorithm using the number of jets capturing the two visible taus as 
the denominator is $7\sim 25$~\% depending on $m_A$, as shown by the dashed curve.
Finally the overall combined tagging efficiency for the signal ranges from 5\% to 0.5\% are shown by the red curve, while the mistagging efficiency for the non-tau jets is $\sim {\cal O} (0.1)$~\%.  Another analysis quotes a similar di-tau tagging/mis-tag efficiency~\cite{Conte:2016zjp}.

The main background after the appropriate preselection would be $t\bar{t}$, as considered in this paper. 
A more dedicated analysis would require collaborations with experimental inputs.  
The set of the preselection is as follows:
\begin{itemize}
\item require that the event contains exactly one isolated lepton $\ell$ and at least three jets, 
with exactly one of them tagged as a $b$-jet and exactly one of them tagged with di-$\tau$, $j_{\tau\tau}$;
\item $m_{b\ell} < 150$~GeV and $m_T(\ell, E\!\!\!/_T)<100$~GeV to guarantee that the event contains one standard top decay; and
\item $100~\gev<m_{j_{\tau\tau} j_1} < 200~\gev$ to make sure $j_{\tau\tau}$ and $j_1$ are from the rare  $t\to u A$ decay, where $j_1$ is the hardest non-$b$, non-di-$\tau$ jet. 
\end{itemize}
The reconstructed $m_{\rm top}^{\rm rec}= m_{j_{\tau\tau} j_1}$ distributions for the signal samples and the $t\bar{t}$ background are shown in the upper right plot of Fig.~\ref{fig:ditau}.  The signal events peak around $150$~GeV.  Finally, to reduce the remaining $t\bar{t}$ background, we make use of the $m_{j_{\tau\tau}}$ distributions shown in the lower left plot of Fig.~\ref{fig:ditau} and apply the $m_{j_{\tau\tau}} > 10$~GeV cut.  The mass of the signal di-$\tau$ jet has a peak slightly below the corresponding $m_A$.  Based on the remaining number of events after all the selection cuts, we estimate and show the sensitivity to the $BR(t\to uA)$ in the lower right plot of Fig.~\ref{fig:ditau}.
The resulting current sensitivity would reach below 0.1~\% for $m_A\sim 20$~GeV, 
which is corresponding to $2\times 10^{-3}$ sensitivity on $\rho_u$ for $\tan\beta=40$,
 and would provide a better sensitivity for $m_A \lesssim 45$~GeV compared with the limit given in Fig~\ref{fig:taupt} in the previous section.
We also show the future prospects of the sensitivity using 300~fb$^{-1}$ (3000~fb$^{-1}$) at 14-TeV LHC.  They would reach ${\cal O} (0.003-0.02)$~\% for $m_A=15 \sim 20$~GeV depending on the systematic uncertainty assumption.  It will be a factor of $4 - 25$ improvement from the current constraints. It would be translated to $4\times 10^{-4} \sim 10^{-3}$ sensitivity on $\rho_u$ for $\tan\beta=40$.
The dotted, dashed and solid curves in the plot correspond to the assumed systematic uncertainties of 
 5~\%, 1~\%, and 0~\%, respectively.

\subsection{Flavor-violating decay of heavy Higgses}
Another smoking gun signature of this model is the flavor-violating decays of heavy Higgs bosons $H$ and $H^\pm$, 
where we mean flavor-violating $H^\pm$ decays as those including different generations in the final states.\footnote{The SM CKM matrix also initiates such modes, but a more significant fraction of the branching ratio would be expected in the VAM's with non-zero $\rho_u$.}
Since the characteristic helicity structure is expected,
a sizable branching ratio of $H \to t u$ (including both $H \to t_L \bar{u}_L$ and $H \to u_R \bar{t}_R$ for short) 
and the corresponding $H^+ \to u_R \bar{b}_R$ should be observed.
Existence of these decays would offer a clear difference between the simple type-X 2HDM and the up-specific VAM.
Note that when $A$ is heavy, though loosing the motivation for explaning $(g-2)_\mu$, 
the corresponding flavor-violating decay modes of $A$ are also predicted. 
A modified model with the capacity to accommodate $(g -2)_\mu$ in the heavy $A$ scenario will be discussed in the next section.

\begin{figure}[h!]
\centering
\includegraphics[height=2.7in]{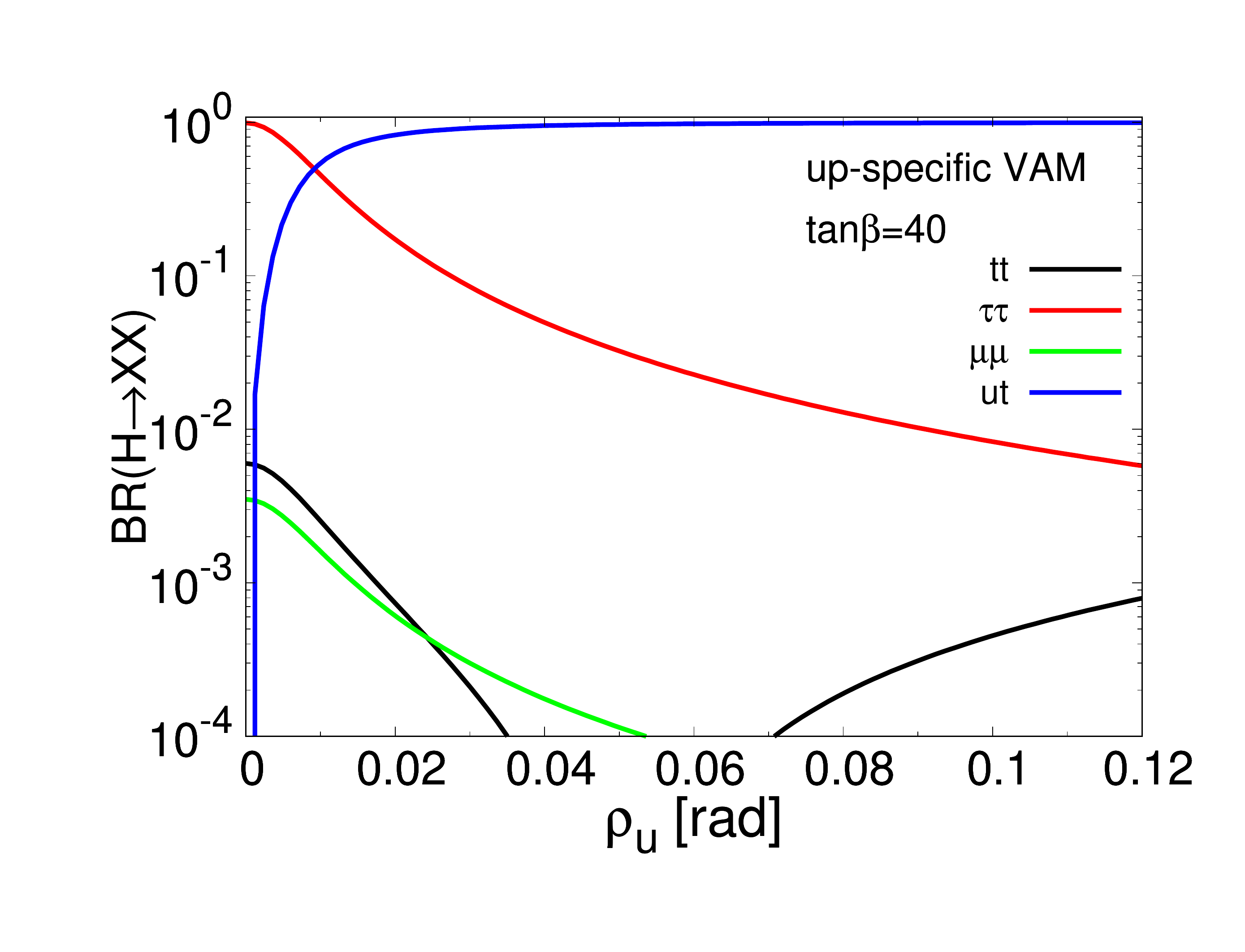}
\caption{The branching ratios of the heavy $H$ decays in the up-specific VAM as a function of $\rho_u$ for $\tan\beta=40$ and $c_{\beta-\alpha} = 0$. Only the fermionic decay modes are considered.
} 
\label{fig:br_upspecific}
\end{figure}

For $m_H \gg m_t$ and $\tan \gg 1$, we have 
\begin{eqnarray}
\frac{BR(H \to t u)}{BR(H \to \tau\tau)}
\sim \frac{m_t^2}{m_\tau^2}\frac{3\sin^2\rho_u}{2} 
\simeq \left( 120 \cdot \sin\rho_u \right)^2
~.
\end{eqnarray}
Therefore, the flavor-violating decay $H \to t u$ dominates for $\rho_u \gtrsim 1/120$.
Fig.~\ref{fig:br_upspecific} shows the branching ratios $BR(H \to ff')$ as a function of $\rho_u$ for 
$\tan\beta=40$ and $c_{\beta-\alpha}=0$.
For example, $BR(H\to tu)$ reaches 90\% for $\rho_u \sim \pi/100$. 
Since $BR(H\to \tau\tau)$ is suppressed due to the new decay mode, 
the constraint from the searches for the heavy Higgs bosons via the $\tau\tau$ mode could become much weaker in our scenario.
It will be even more suppressed with the existence of the other decay modes involving the Higgs bosons, although we assume them negligible here.

Our model also predicts the helicities of top quark in the decay products from the heavy Higgses:
the left-handed top or the right-handed anti-top should be observed.  Confirming the existence of the $H\to tu$ decay and measuring the polarization of $t$ in the decay products 
would be an important test for our model.

\section{Another variants}
\label{sec:down_type_variant}
\subsection{Muon-specific in lepton sector}
So far, we have assigned the same non-zero PQ charge to charged leptons of all three generations,
making their Yukawa couplings to $A$ enhanced by $\tan\beta$.
In particular, the enhanced $A\tau\tau$ coupling is important to enhance the 
Barr-Zee diagram contributions to $(g-2)_\mu$ while it is also constrained by 
the lepton universalities in heavy lepton and $Z$ decays.
Thus, the mass of $A$ is required to be lighter than 30~GeV as shown in Fig.~\ref{fig:g-2}, 
which forces us to fine-tune the $hAA$ coupling to zero 
as $h\to AA$ decay is always kinematically allowed.
For the lepton sector, however, we have more freedom in the PQ charge assignment 
as they are not relevant to the PQ-solution nor the domain wall problem.  From the phenomenological 
point of view, we can assign a non-zero PQ charge only to $\mu$ and keep $\tau$ and $e$ PQ-neutral. 
We shall refer to the lepton sector in this scenario as the muon-specific lepton sector.  
It might be even more natural as we have a parallel setup to treat only one generation being special in both lepton and quark sectors. 
As a result, the $A\tau\tau$ and $H\tau\tau$ couplings are suppressed by $\cot\beta$, and the constraints 
from the heavy lepton and the $Z$ decays become irrelevant. 
In this case, the 1-loop contribution dominates over the 2-loop contribution of the 
$\cot\beta$ suppressed $\tau$-loop Barr-Zee diagram for $(g-2)_{\mu}$~\cite{Abe:2017jqo}.
Since among the 1-loop contributions only the one involving the CP-even $H$ provides a positive contribution to $(g-2)_{\mu}$, as explicitly seen in Table~\ref{tab:g-2contribution}, 
the preferred parameter region has to satisfy $m_H<m_A$. 
As all the contributions roughly scale like $\tan^2 \beta/m_\phi^2$, the required value of $\tan \beta/m_\phi$ to explain the deviation in $(g-2)_\mu$ is about $7 \cdot 10^{3}~\tev^{-1}$.

In the simple muon-specific model, 
the LHC data already set lower bounds on $m_H$ and $m_A$ as a plethora of $4\mu$ events 
would be generated through $pp\to Z^* \to HA$ and both $H$ and $A$ can decay 
into a pair of muons with a branching ratio of almost 100\% due to the $\tan\beta$ enhancement.
The search with three or more muons had been performed by the CMS Collaboration
using the 13-TeV data with 35.9~fb$^{-1}$ and found the data consistent with the SM expectation~\cite{cms-multi}, 
which constrains $m_H\gtrsim 640$~GeV when we assume $m_H \sim m_A$~\cite{Abe:2017jqo}.
In this case, the $h \to AA$ decay is kinematically forbidden, and we do not have to worry about the fine-tuning problem of the $hAA$ coupling.
With such heavy Higgs masses, $\tan\beta \sim 3000$ would be required to explain the $(g-2)_\mu$ deviation.

With the muon-specific lepton sector, only the up-specific VAM would be valid among the up-type VAM's 
as the $\tan\beta$ enhanced 2-loop contributions to $(g-2)_\mu$ are negative for the up-type quarks and 
only the up Yukawa coupling is small enough to neglect the effects.
As the mixing controlled by $\rho_u$ initiates the $\tan\beta$ enhanced 2-loop $t$-contribution, we cannot consider arbitrary large $\rho_u$, as it is not theoretically favored by perturbativity with such a large $\tan\beta$.
We restrict the range of $\rho_u$ by requiring all Yukawa couplings be perturbative, and obtain 
$\rho_u \lesssim 20/\tan\beta \sim 0.006$  from $Y_{ut}^\prime=\sqrt{2}\xi_{ut}m_t/v \lesssim 4\pi$. 
Within this range, the negative 2-loop top contribution is negligible.

\begin{figure}[h!]
\centering
\includegraphics[height=2.7in]{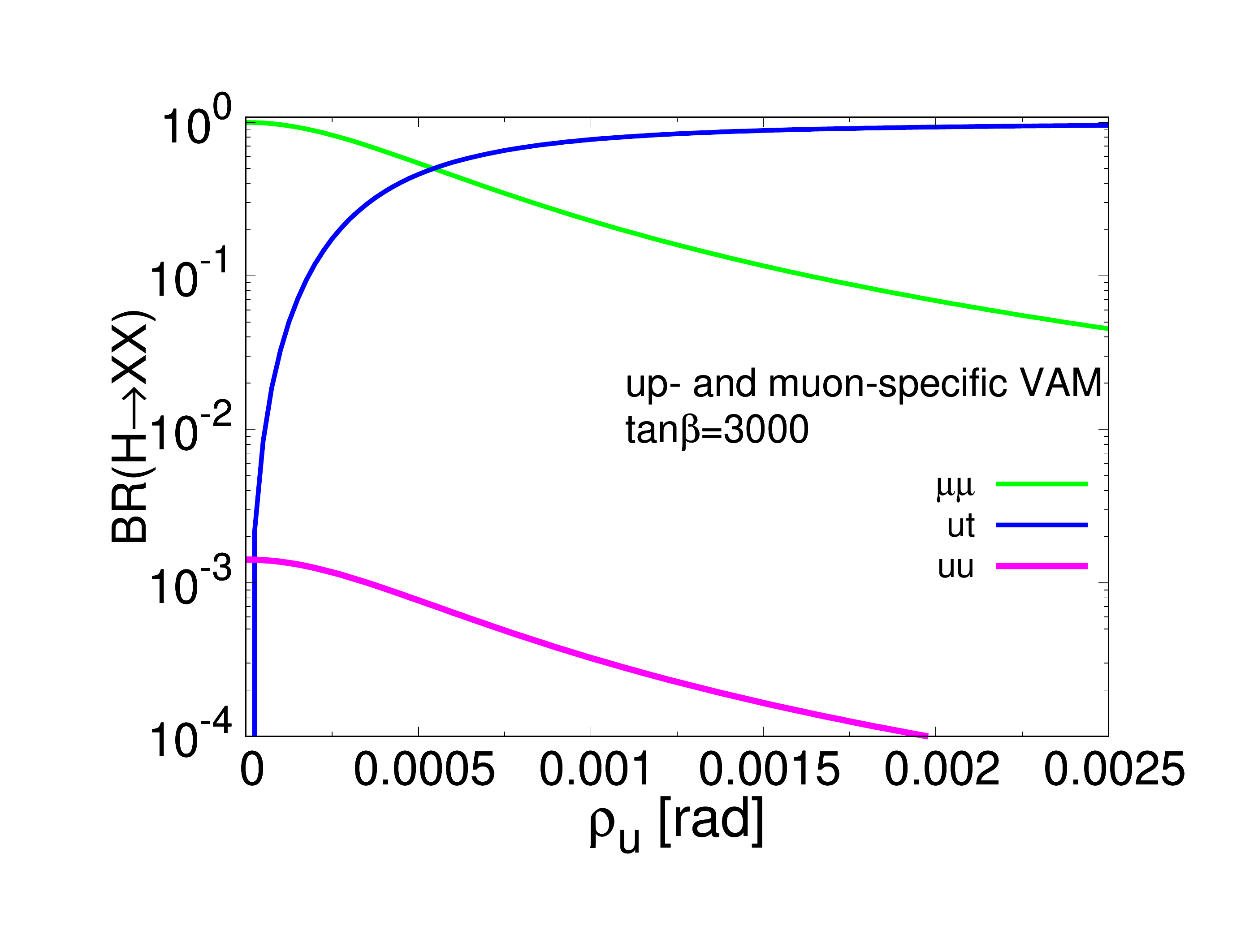}
\caption{Branching ratios of the heavy Higgs boson $H$ in the up-specific VAM with the muon-specific lepton sector as a function of $\rho_u$ for $\tan\beta=3000$ and $c_{\beta-\alpha} = 0$. Only the fermionic decay modes are considered.}
\label{fig:br_upmuspecific}
\end{figure}

As $\rho_u$ increases, the enhanced branching ratio of $A/H \to tu$ reduces and dominates over 
the decay of $A/H \to \mu\mu$ for a good portion of the parameter space.
Explicitly, the ratio of their branching ratios is
\begin{eqnarray}
\frac{BR(A \to t u)}{BR(A \to \mu\mu)}
\simeq 
\frac{BR(H \to t u)}{BR(H \to \mu\mu)}
\simeq \frac{m_t^2}{m_\mu^2}\frac{3 \sin^2\rho_u}{2} 
\simeq \left( 2000 \cdot \rho_u \right)^2
\end{eqnarray}
and ranges from 0 to ${\cal O}(10^2)$.
In this case, the above-mentioned $4\mu$ constraint becomes weaker or invalid for a non-zero $\rho_u$. 
It opens up an allowed region for lighter $m_A$ with the corresponding smaller $\tan\beta$.
Fig.~\ref{fig:br_upmuspecific} shows the branching fractions 
$BR(H \to ff')$ as a function of $\rho_u$ assuming only the fermionic decay modes contribute.
Observing the flavor-violating $A/H \to tu$ decays would be a smoking gun signature to distinguish 
between the simple muon-specific 2HDM and the up-type VAM with the muon-specific lepton sector.  
In this scenario, the flavor-changing rare top decays, $t \to u H $ and $t \to u A $, 
are kinematically forbidden.

As mentioned above, the charm-specific VAM is difficult to accommodate such large $\tan\beta$ 
because the 2-loop contribution would dominate over the 1-loop contributions, resulting in an opposite contribution to the $(g-2)_\mu$.
The top-specific VAM is also disfavored by the same reason as the perturbativity of Yukawa couplings.
Among the down-type specific VAM's, down-specific and strange-specific VAM's would be compatible with the muon-specific lepton sector as the corresponding Yukawa couplings are negligible, analogous to the up-specific case.
In the bottom specific VAM, since the $\tan^2\beta$ enhanced bottom contribution to $(g-2)_\mu$
dominates over the 1-loop contribution, $m_A < m_H$ is favored and $\tan\beta/m_A \sim 10^3~\tev^{-1}$ is required. 
In this setup, the dominant decay mode of $H$ and $A$ becomes $bb$ and 
the $\mu\mu$ mode is suppressed by $(m_\mu/m_b)^2$, which makes 
the above-mentioned $4\mu$ constraint weaker.  
However, such an enhancement in the bottom Yukawa coupling 
would require an extreme fine-tuning at the level of $\tan^{-4}\beta$ to accommodate 
the $B_s\to \mu\mu$ decay branching ratio, since the bottom diagrams contribute 
$\Delta P ={\cal O}(0.1) \tan^4\beta$ as discussed in Sec.~\ref{sec:current_bounds}.
\bigskip

\subsection{Assigning non-zero PQ charges to down-type RH quarks \label{sec:down-vam}}

Instead of assigning non-zero PQ charge to the RH up-type quarks, we can do the same to the RH down-type quarks without loosing the motivations.  In this subsection, we briefly comment on how such a scenario is severely constrained by quark mixing in the down sector.  The mixing structure is analogous to the up-type specific VAM, and the mixing matrix $V_d$ would be the same in form as $V_u$ but with $\rho_u$ and $\psi_u$ replaced respectively by $\rho_d$ and $\psi_d$ to describe the $d-b$ and $d-s$ mixing.
We also define $H_d$ in a way analogous to $H_u$ in Eq.~(\ref{eq6}) with the corresponding substitution.
According to Table~II of Ref.~\cite{Harnik:2012pb}, the constraints from $B^0_d$, $B^0_s$ and $K^0$ oscillations are: 
\begin{eqnarray}
B^0_d &:&
\frac{|\xi^A_{db}|^2}{2m^2_A}\frac{m^2_b}{v^2} \lesssim  7.4\times 10^{-13}~\gev^{-2}
\Rightarrow
(\tan\beta+\cot\beta) \frac{|H^{db}_d|}{m_A} 
\lesssim  7.2\times 10^{-5}~\gev^{-1}\,,\\
B^0_s &:&
\frac{|\xi^A_{sb}|^2}{2m^2_A}\frac{m^2_b}{v^2} \lesssim  5.8\times 10^{-11}~\gev^{-2}
\Rightarrow
(\tan\beta+\cot\beta) \frac{|H^{sb}_d|}{m_A} 
\lesssim  6.3\times 10^{-4}~\gev^{-1}\,,\\
K^0 &:&
\frac{|\xi^A_{ds}|^2}{2m^2_A}\frac{m^2_s}{v^2} \lesssim  1.8\times 10^{-14}~\gev^{-2}
\Rightarrow
(\tan\beta+\cot\beta) \frac{|H^{ds}_d|}{m_A} 
\lesssim  4.9\times 10^{-4}~\gev^{-1}\,.
\end{eqnarray}

\begin{figure}[h!]
\centering
\includegraphics[height=3in]{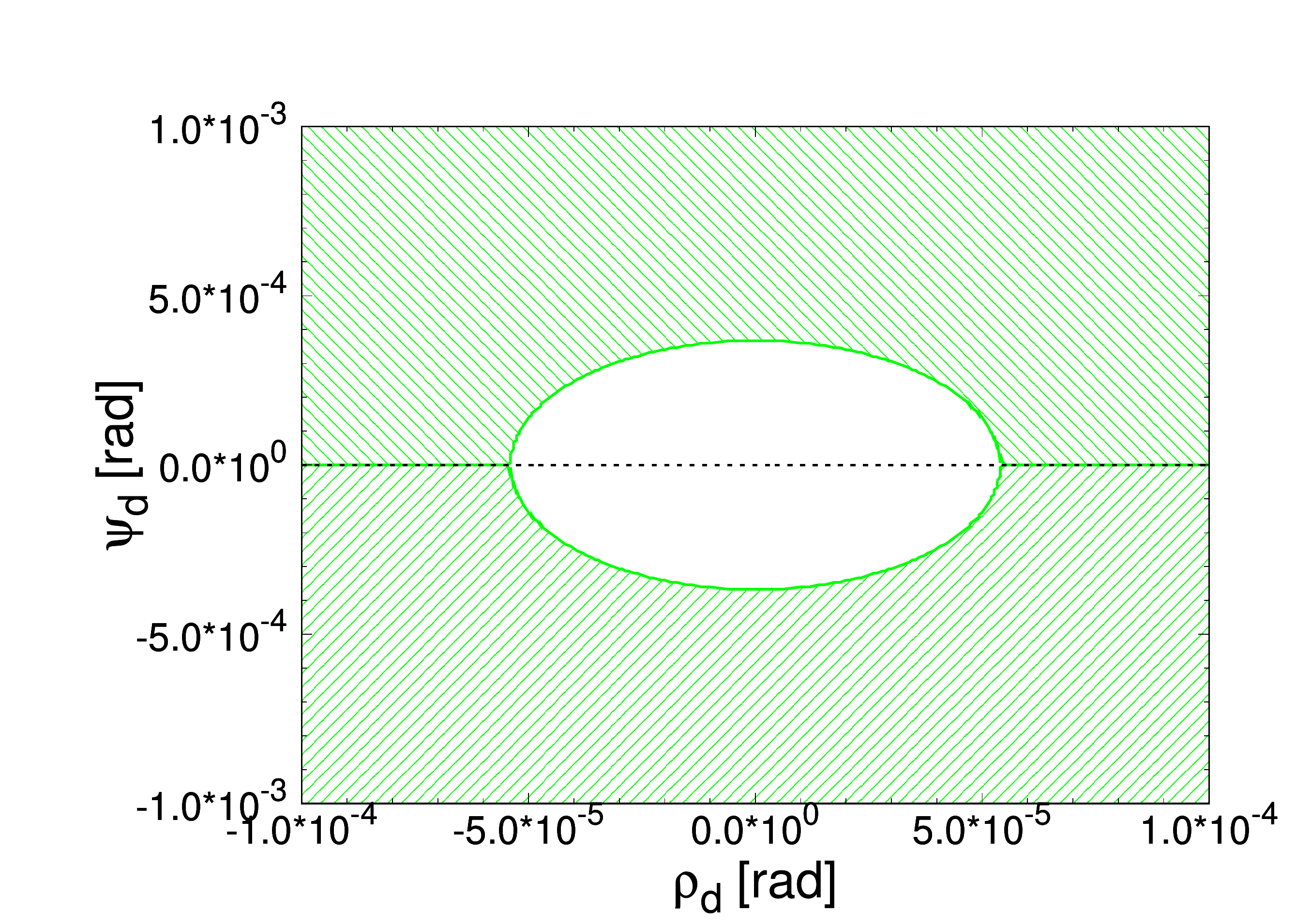}
\caption{\small \label{fig:KK} 
Allowed parameter space in the $\rho_d$-$\psi_d$ plane for the scenario where only the down-type RH quarks in the quark sector carry non-zero PQ charges. Only the region nearby down-specific VAM is shown. 
The green-hatched region is ruled out by combination of the constraints from $B^0_d$, $B^0_s$, and $K^0$ oscillations. 
In this plot, we fix $\tan\beta=40$ and $m_A=15$~GeV.
}
\end{figure}

The region in the $\rho_d-\psi_d$ plane allowed by the above constraints 
for $m_A=15~\gev$ and $\tan\beta=40$ is shown by the white region in Fig.~\ref{fig:KK}.
Only the region nearby $(\rho_d, \psi_d) \sim (0, 0)$ (down-specific VAM) is shown. 
The $B_d^0$ constraint is important for $\psi_d = 0$, giving $\rho_d,\rho_d' \ \lesssim 1.4 \times (m_A/10^4 \gev)/\tan\beta$.
The $K^0$ constraint is important for $\rho_d=0$, giving $\psi_d,\psi_d' \lesssim 9.8 \times (m_A/10^4 \gev)/\tan\beta$.
The $B_s^0$ constraint is important for $\psi_d=\pi$, giving $\rho_d,\rho_d' \lesssim 12.6 \times (m_A/10^4 \gev)/\tan\beta$.
We define $\psi_d = \psi_d^\prime + \pi$ and $\rho_d = \rho_d^\prime + \pi$. 
There are also allowed regions with $(\rho_d, \psi_d) \sim (0, \pi)$ (strange-specific VAM) and $(\pi,\mbox{any value})$ (bottom-specific VAM).
However, as mentioned in the previous section the bottom-specific solution is strongly disfavored 
by the $B_s\to \mu\mu$ data.

\section{conclusions}
\label{sec:conclusion}

We have studied variant axion models (VAM's) with only a specific fermion charged under the Peccei-Quinn symmetry and their capacity to accommodate the muon $g-2$ anomaly as well as the compatibility with various other experimental constraints.  We start by considering the up-type specific VAM's
and find that the combined $\chi^2$ fit favors the parameters $m_A\sim 15~\gev$ and $\tan\beta \sim 40$, the same as the type-X 2HDM. 
Moreover, we find that this parameter choice has no conflict with flavor observables as long as the mixing angle $\rho_u$ is sufficiently small.  In particular, a small nonzero mixing angle $\rho_u \sim \pi/100$ is slightly favored by the observed $B_s \to \mu\mu$ branching ratio.

As the charm-mediated Barr-Zee diagram contribution to $(g-2)_\mu$ is negative, 
the charm-specific VAM is disfavored in comparison with the up-specific VAM.  
We therefore focus on the up-specific model and its promising signature of the 
rare $t\to uA$ decay followed by $A \to \tau\tau$ at the LHC. 
Current searches of $A$ already impose some constraints in the parameter space, 
but do not exclude the most interesting region of $m_A \sim 20~\gev$.  We propose 
an efficient search strategy that employs di-tau tagging using jet substructure information, and 
have explicitly demonstrated that it would enhance the sensitivity on $BR(t \to uA)$, especially 
in the light $m_A$ region of great interest to us.  Our model also predicts that the heavy Higgs bosons have significant flavor-violating decays, such as $A/H \to t u$.
We encourage our experimental colleagues to search intensively for this flavor-changing top decay and the flavor-violating resonances.  

We have also considered other variants: the muon-specific lepton sector and the down-type specific VAM's. 
The up-specific VAM with the muon-specific lepton sector is very interesting possibility
as no tuning is required to suppress $h \to AA$ and the scenario is not constrained by the lepton universality measurements.
Unlike the simplest muon-specific model, the up-specific VAM with the muon-specific sector 
predicts that the heavy Higgs bosons can decay into a pair of flavor-violating up-type quarks 
such as $H/A \to t u$ at a significant branching fraction.
It suppresses the $H/A \to \mu\mu$ decay, making the $4\mu$ constraint at the LHC less effective and opens up more parameter space.
The down/strange-specific VAM's with the muon-specific lepton sector would also be viable possibilities.
The down-type specific VAM's are strongly constrained by the $B_s\to \mu^+\mu^-$ decay and the $B_{d,s}$ and $K$ meson mixing data, rendering a very fine-tuned parameter space.  Nevertheless, such scenarios could offer another interesting possibility to explain $(g-2)_\mu$ as one of the bottom 
Barr-Zee diagram contribution is positive.
%

\section*{Acknowledgments}

The authors would like to thank Kei Yagyu for some discussions about the $muon$-specific lepton sector.
This research was supported in part by the Ministry of Science and Technology of Taiwan under Grant No.\ NSC 100-2628-M-008-003-MY4 (C.-W.~C); 
the Japan Society for the Promotion of Science (JSPS)  Grant-in-Aid for Scientific Research Numbers No.~26104001, No.~26104009, No.~16H02176, and No.~17H02878 (T.~T.~Y.); 
and the World Premier International Research Center Initiative, MEXT, Japan (M.~T., P.-Y.~T, and T.~T.~Y.). 
MT is supported in part by the JSPS Grant-in-Aid for Scientific Research Numbers~16H03991,~16H02176,~17H05399, and~18K03611.


\begin{thebibliography}{99}

\bibitem{Peccei:1977hh} 
R.~D.~Peccei and H.~R.~Quinn,
Phys.\ Rev.\ Lett.\  {\bf 38}, 1440 (1977).

\bibitem{Weinberg:1977ma} 
S.~Weinberg,
Phys.\ Rev.\ Lett.\  {\bf 40}, 223 (1978).

\bibitem{Wilczek:1977pj} 
F.~Wilczek,
Phys.\ Rev.\ Lett.\  {\bf 40}, 279 (1978).


\bibitem{Agashe:2014kda} 
K.~A.~Olive {\it et al.}  [Particle Data Group Collaboration],
Chin.\ Phys.\ C {\bf 38}, 090001 (2014).

\bibitem{Abbott:1982af} 
L.~F.~Abbott and P.~Sikivie,
Phys.\ Lett.\ B {\bf 120}, 133 (1983).

\bibitem{Preskill:1982cy} 
J.~Preskill, M.~B.~Wise and F.~Wilczek,
Phys.\ Lett.\ B {\bf 120}, 127 (1983).

\bibitem{Dine:1982ah} 
M.~Dine and W.~Fischler,
Phys.\ Lett.\ B {\bf 120}, 137 (1983).

\bibitem{Ade:2015xua} 
P.~A.~R.~Ade {\it et al.}  [Planck Collaboration],
arXiv:1502.01589 [astro-ph.CO].

\bibitem{Peccei:1986pn} 
R.~D.~Peccei, T.~T.~Wu and T.~Yanagida,
Phys.\ Lett.\ B {\bf 172}, 435 (1986).

\bibitem{Krauss:1986wx} 
L.~M.~Krauss and F.~Wilczek,
Phys.\ Lett.\ B {\bf 173}, 189 (1986).

\bibitem{Chen:2010su} 
C.~R.~Chen, P.~H.~Frampton, F.~Takahashi and T.~T.~Yanagida,
JHEP {\bf 1006}, 059 (2010)
[arXiv:1005.1185 [hep-ph]].

\bibitem{Chiang:2015cba} 
  C.~W.~Chiang, H.~Fukuda, M.~Takeuchi and T.~T.~Yanagida,
  JHEP {\bf 1511}, 057 (2015)
  [arXiv:1507.04354 [hep-ph]].

\bibitem{Chiang:2017fjr} 
  C.~W.~Chiang, H.~Fukuda, M.~Takeuchi and T.~T.~Yanagida,
  Phys.\ Rev.\ D {\bf 97}, no. 3, 035015 (2018)
  [arXiv:1711.02993 [hep-ph]].

\bibitem{Hagiwara:2011af} 
  K.~Hagiwara, R.~Liao, A.~D.~Martin, D.~Nomura and T.~Teubner,
  J.\ Phys.\ G {\bf 38}, 085003 (2011)
  [arXiv:1105.3149 [hep-ph]].

\bibitem{2HDMg2} 
  E.~J.~Chun,
  EPJ Web Conf.\  {\bf 118}, 01006 (2016)
  [Pramana {\bf 87}, no. 3, 41 (2016)]
  [arXiv:1511.05225 [hep-ph]].
  
\bibitem{Chun:2015hsa} 
  E.~J.~Chun, Z.~Kang, M.~Takeuchi and Y.~L.~S.~Tsai,
  JHEP {\bf 1511}, 099 (2015),
  [arXiv:1507.08067 [hep-ph]].
  
\bibitem{Abe:2017jqo} 
  T.~Abe, R.~Sato and K.~Yagyu,
  JHEP {\bf 1707}, 012 (2017),
  [arXiv:1705.01469 [hep-ph]].
  
\bibitem{Bennett:2006fi} 
  G.~W.~Bennett {\it et al.} [Muon g-2 Collaboration],
  Phys.\ Rev.\ D {\bf 73}, 072003 (2006),
  [hep-ex/0602035].
  
  
\bibitem{Jegerlehner:2009ry} 
  F.~Jegerlehner and A.~Nyffeler,
  Phys.\ Rept.\  {\bf 477}, 1 (2009),
  [arXiv:0902.3360 [hep-ph]].
  
\bibitem{Broggio:2014mna} 
  A.~Broggio, E.~J.~Chun, M.~Passera, K.~M.~Patel and S.~K.~Vempati,
  JHEP {\bf 1411}, 058 (2014),
  [arXiv:1409.3199 [hep-ph]].
 
\bibitem{hfag} 
  Y.~Amhis {\it et al.} [Heavy Flavor Averaging Group (HFAG)],
  [arXiv:1412.7515 [hep-ex]].

\bibitem{ALEPH:2005ab} 
  S.~Schael {\it et al.} [ALEPH and DELPHI and L3 and OPAL and SLD Collaborations and LEP Electroweak Working Group and SLD Electroweak Group and SLD Heavy Flavour Group],
  Phys.\ Rept.\  {\bf 427}, 257 (2006),
  [hep-ex/0509008].
    
\bibitem{zdecay} 
  E.~J.~Chun and J.~Kim,
  JHEP {\bf 1607}, 110 (2016),
  [arXiv:1605.06298 [hep-ph]].
  
\bibitem{bs-LHCb} 
  R.~Aaij {\it et al.} [LHCb Collaboration],
  Phys.\ Rev.\ Lett.\  {\bf 111}, 101805 (2013),
  [arXiv:1307.5024 [hep-ex]].
  
\bibitem{bs-CMS} 
  S.~Chatrchyan {\it et al.} [CMS Collaboration],
  Phys.\ Rev.\ Lett.\  {\bf 111}, 101804 (2013),
  [arXiv:1307.5025 [hep-ex]].
  
\bibitem{bs} 
  X.~Q.~Li, J.~Lu and A.~Pich,
  JHEP {\bf 1406}, 022 (2014),
  [arXiv:1404.5865 [hep-ph]].

\bibitem{Harnik:2012pb} 
  R.~Harnik, J.~Kopp and J.~Zupan,
  JHEP {\bf 1303}, 026 (2013),
  [arXiv:1209.1397 [hep-ph]].
    
\bibitem{CMS:2016hdd} 
  CMS Collaboration [CMS Collaboration],
  CMS-PAS-TOP-16-019.

\bibitem{Sirunyan:2018pzn}
 A.~M.~Sirunyan {\it et al.} [CMS Collaboration],
 arXiv:1805.10191 [hep-ex].
 
\bibitem{Curtin:2013fra} 
  D.~Curtin {\it et al.},
  Phys.\ Rev.\ D {\bf 90}, no. 7, 075004 (2014),
  [arXiv:1312.4992 [hep-ph]].

\bibitem{dimuon} 
  I.~Hoenig, G.~Samach and D.~Tucker-Smith,
  Phys.\ Rev.\ D {\bf 90}, no. 7, 075016 (2014),
  [arXiv:1408.1075 [hep-ph]].

  
\bibitem{Kao:2011aa}
 C.~Kao, H.~Y.~Cheng, W.~S.~Hou and J.~Sayre,
 Phys.\ Lett.\ B {\bf 716}, 225 (2012)
 [arXiv:1112.1707 [hep-ph]].

\bibitem{Chen:2013qta}
 K.~F.~Chen, W.~S.~Hou, C.~Kao and M.~Kohda,
 Phys.\ Lett.\ B {\bf 725}, 378 (2013)
 [arXiv:1304.8037 [hep-ph]].


\bibitem{Altunkaynak:2015twa}
 B.~Altunkaynak, W.~S.~Hou, C.~Kao, M.~Kohda and B.~McCoy,
 Phys.\ Lett.\ B {\bf 751}, 135 (2015)
 [arXiv:1506.00651 [hep-ph]].
  
\bibitem{CMS:2015mca} 
  CMS Collaboration [CMS Collaboration],
  CMS-PAS-HIG-14-029.
  
\bibitem{Khachatryan:2015baw} 
  V.~Khachatryan {\it et al.} [CMS Collaboration],
  Phys.\ Lett.\ B {\bf 758}, 296 (2016)
  [arXiv:1511.03610 [hep-ex]].

\bibitem{Sirunyan:2017uvf} 
  A.~M.~Sirunyan {\it et al.} [CMS Collaboration],
  JHEP {\bf 1711}, 010 (2017)
  [arXiv:1707.07283 [hep-ex]].

\bibitem{Goncalves:2016qhh} 
  D.~Goncalves and D.~Lopez-Val,
  Phys.\ Rev.\ D {\bf 94}, no. 9, 095005 (2016)
  [arXiv:1607.08614 [hep-ph]].
  
\bibitem{Banerjee:2017wxi} 
  S.~Banerjee, D.~Barducci, G.~Belanger, B.~Fuks, A.~Goudelis and B.~Zaldivar,
  JHEP {\bf 1707}, 080 (2017)
  [arXiv:1705.02327 [hep-ph]].
  
\bibitem{Bernon:2014nxa} 
  J.~Bernon, J.~F.~Gunion, Y.~Jiang and S.~Kraml,
  Phys.\ Rev.\ D {\bf 91}, no. 7, 075019 (2015)
  [arXiv:1412.3385 [hep-ph]].
  
\bibitem{Aaboud:2017sjh} 
  M.~Aaboud {\it et al.} [ATLAS Collaboration],
  JHEP {\bf 1801}, 055 (2018)
  [arXiv:1709.07242 [hep-ex]].
  
\bibitem{Alwall:2014hca} 
  J.~Alwall {\it et al.},
  JHEP {\bf 1407}, 079 (2014)
  [arXiv:1405.0301 [hep-ph]].
  
\bibitem{Sjostrand:2007gs} 
  T.~Sjostrand, S.~Mrenna and P.~Z.~Skands,
  Comput.\ Phys.\ Commun.\  {\bf 178}, 852 (2008)
  [arXiv:0710.3820 [hep-ph]].

\bibitem{delphes3}
  J.~de Favereau {\it et al.} [DELPHES 3 Collaboration],
  JHEP {\bf 1402}, 057 (2014)

\bibitem{Cacciari:2011ma} 
  M.~Cacciari, G.~P.~Salam and G.~Soyez,
  Eur.\ Phys.\ J.\ C {\bf 72}, 1896 (2012)
  [arXiv:1111.6097 [hep-ph]].

\bibitem{Papaefstathiou:2014oja} 
  A.~Papaefstathiou, K.~Sakurai and M.~Takeuchi,
  JHEP {\bf 1408}, 176 (2014)
  [arXiv:1404.1077 [hep-ph]].

\bibitem{Katz:2010iq} 
  A.~Katz, M.~Son and B.~Tweedie,
  Phys.\ Rev.\ D {\bf 83}, 114033 (2011)
  [arXiv:1011.4523 [hep-ph]].
  
\bibitem{Conte:2016zjp} 
  E.~Conte, B.~Fuks, J.~Guo, J.~Li and A.~G.~Williams,
  JHEP {\bf 1605}, 100 (2016)
  [arXiv:1604.05394 [hep-ph]].

\bibitem{cms-multi} 
  CMS Collaboration [CMS Collaboration], CMS-PAS-EXO-17-006,
  [arXiv:1708.07962 [hep-ex]].
  
  
\end{thebibliography}
\end{document}